\documentclass[12pt]{article}
\usepackage{graphicx}
\usepackage{axodraw}
\usepackage{amsmath}

\makeatletter
  \newcommand{\preprintnumber}{---DRAFT---\\\today}
  \newcommand{\ps@preprinttitle}{%
    \let\@oddfoot\@empty
    \let\@evenfoot\@empty
    \def\@oddhead{\parbox[t]{\textwidth}{%
                    \begin{flushright}%
                      \preprintnumber
                    \end{flushright}}}
    \let\@evenhead\@oddhead}
\makeatother
\begin{document}
\title{%
  Complete Calculations of $Wb\bar{b}$ and $Wb\bar{b}+\text{jet}$
  Production at Tevatron and LHC:\\
  Probing Anomalous $Wtb$ Couplings in Single~Top~Production}
\author{%
  E. Boos$^1$, L. Dudko$^1$ and T. Ohl$^2$\\
  \hfil\\
  $^1$Institute of Nuclear Physics, Moscow State University,\\
      119899, Moscow, Russia \\
  $^2$Institut f\"ur Kernphysik, Darmstadt University of Technology,\\
      Darmstadt, Germany}
\renewcommand{\today}{\relax}
\maketitle
\begin{abstract}
  We present the results of a complete tree level calculation of the
  processes $pp(\bar{p}) \to Wb\bar{b}$ and
  $Wb\bar{b}+\text{jet}$ that includes the single top signal and all
  irreducible backgrounds simultaneously. In order to probe the
  structure of the $Wtb$ coupling with the highest possible accuracy
  and to look for possible deviations from standard model predictions,
  we identify sensitive observables and propose an optimal set of cuts
  which minimizes the background compared to the signal. At the LHC,
  the single top and the single anti-top rates are different and the
  corresponding asymmetry yields additional information. The analysis
  shows that the sensitivity for anomalous couplings will be improved
  at the LHC by a factor of 2--3 compared to the expectations for the
  first measurements at the upgraded Tevatron. Still, the bounds on
  anomalous couplings obtained at hadron colliders will remain 2--8
  times larger than those from high energy $\gamma e$ colliders, which
  will, however, not be available for some time.  All basic calculations
  have been carried out using the computer package Comp\-HEP. The known
  NLO corrections to the single top rate have been taken into account.
\end{abstract}
\thispagestyle{preprinttitle}
\renewcommand{\preprintnumber}{%
   INP-MSU 99-4/562\\
   IKDA-99/02\\
   hep-ph/9903215\\
   March 1, 1999}
\setcounter{page}{0}
\newpage

%%%%%%%%%%%%%%%%%%%%%%%%%%%%%%%%%%%%%%%%%%%%%%%%%%%%%%%%%%%%%%%%%%%%%%%%
\section{Introduction}
The observation by the CDF and D0 collaborations~\cite{cdfd0} of a
very heavy top quark with a mass of about~175\,GeV, close to the indirect
prediction from fits of precision electroweak
data~$177^{+7+16}_{-7-19}$\,GeV~\cite{ew}, has been an important 
confirmation of the Standard Model~(SM).  Still, important open
problems remain: why is the top is so heavy and is it really a point
like particle?  A curious numerical coincidence between the mass and
vacuum expectation value~$v/\sqrt{2}=175$\,GeV sets the top quark
Yukawa coupling close to unity.  As has been stressed some time
ago~\cite{peccei}, because of such unique properties, the top quark
might provide for the first time a window to the physics of
electroweak symmetry breaking and even ``new physics'' beyond.
 
A possible signal would be deviations from the SM predictions of the
interactions of the top quark with other fields.  Therefore it is
important to study and measure all top quark couplings, in particular
the coupling to the $W$-boson and the $b$-quark which is responsible
for almost all top quark decays.
Therefore, events with the production of a single top quark are
extremely interesting at different colliders, because they are
directly proportional to the $Wtb$~vertex.  Thus one can hope to
measure the structure of the vertex and possible deviations from the
SM predictions with a high accuracy.  Furthermore, it should be noted
that the processes of single top production and decay involve light
Higgs production simultaneously (cf.~the discussion in~\cite{boos1}).
The measurement of the $Wtb$ vertex in $\gamma e$ collisions has been
described in~\cite{boos1, young}.
 
In this paper we discuss the possible accuracy in the determination of
the structure of the $Wtb$ vertex at the upgraded Tevatron collider
and at the LHC.  Improving on previous considerations~\cite{previous},
we perform a complete tree level calculation, taking into account
contributions from anomalous operators to the $Wbt$ vertex, the
production of $Wb\bar{b}$ and $Wb\bar{b}+\text{jet}$, which includes
the single top signal together with the irreducible backgrounds.  We
have included the NLO corrections to the single top
part~\cite{willenbrock}.  Based on an analysis of the singularities of
Feynman diagrams and on explicit calculations we identify the set of
the most sensitive variables and their corresponding optimal cuts.
This allows us to obtain a clean single top sample above the
background and a handle on possible deviations from the SM
expectations.

%%%%%%%%%%%%%%%%%%%%%%%%%%%%%%%%%%%%%%%%%%%%%%%%%%%%%%%%%%%%%%%%%%%%%%%%
\section{The Basic Processes}
Single top production at hadron colliders has been studied by a number
of authors (cf.~\cite{previous, willenbrock, boos2} and references
therein). So far, the most complete set of SM processes contributing
to the single top rate has been studied in~\cite{boos2} and the most
accurate NLO calculations to the main processes have been presented
in~\cite{willenbrock}. In recent papers~\cite{boos3, willenbrock1}, 
complete Monte Carlo~(MC) analysises of the single top signal versus
backgrounds have been presented.  The Feynman diagrams for all
processes contributing to the single top production rate have been
presented previously (cf.~e.\,g.~\cite{boos3}) and include virtual~$W$
$s$-channel exchange, $W$-gluon fusion and $W+\text{top}$
production. The first process is the simplest
$2\to2$~reaction, while the $W$-gluon process includes
$2\to3$~parton diagrams.  In order to resum the large QCD
corrections from $g\to b\bar{b}$ splitting in the latter, they
are combined with the $2\to2$ process involving a $b$ quark in
the initial state~\cite{previous, willenbrock, boos2} and the
corresponding piece of the $g\to b\bar{b}$ splitting function
is subtracted in order to avoid double counting.

Finally, the $W+\text{top}$ production process gives a large
contribution to total single top production at the LHC~\cite{boos3}.
However we will not consider this process here, because it does not
contribute to the topologies which we will be interested in.  Moreover,
it has a final state similar to top pair production and after a suitable
background subtraction it is less sensitive to the $Wtb$~vertex
(cf.~the discussion of the corresponding process in
$e^+e^-$~collisions~\cite{jikia-boos}).

Irrespective of the different strategies employed, the analysises
of~\cite{boos3} and~\cite{willenbrock1} have both shown that the
single top production rate is large enough to be visible above the
backgrounds using proper cuts.  This is possible despite the fact that
the background reduction is much more complicated compared to the case
of top pair production.

However, in order to probe the $Wtb$~vertex, one must check for
possible deviations from the SM predictions and then all SM
contributions become part of the background.  Therefore it is
necessary to find even stronger cuts, defining phase space regions
where the deviations from the SM predictions for single top production
will be most prominent.  Approaching this problem we perform
an accurate calculation of the two processes
\begin{subequations}
\label{eq:pp->bbW(j)}
\begin{align}
\label{eq:pp->bbW}
  pp &\to b\bar{b}W  \\
\intertext{and}
\label{eq:pp->bbWj}
  pp &\to b\bar{b}W+\text{jet}
\end{align}
\end{subequations}
which include simultaneously the single top signal and the irreducible
backgrounds.
\begin{figure}
  \begin{center}
% CompHEP  version  3.2    
% diagrams for process   PROCESS:  u,D -> b,B,W+                 15 diagrams
{\def\chepscale{1.3} % picture size control
\unitlength=\chepscale pt
\SetWidth{0.7}      % line    size control
\SetScale{\chepscale}
\scriptsize    %  letter  size control
%  diagram # 1
\begin{picture}(70,60)(0,0)
\Text(15.9,39.8)[r]{$u$}
\ArrowLine(16.3,39.8)(29.0,39.8) 
\Text(35.1,40.4)[b]{$A$}
\DashLine(29.0,39.8)(41.7,39.8){3.0} 
\Text(54.8,49.3)[l]{$b$}
\ArrowLine(41.7,39.8)(54.3,49.3) 
\Text(54.8,30.3)[l]{$B$}
\ArrowLine(54.3,30.3)(41.7,39.8) 
\Text(27.4,30.3)[r]{$u$}
\ArrowLine(29.0,39.8)(29.0,20.8) 
\Text(15.9,20.8)[r]{$D$}
\ArrowLine(29.0,20.8)(16.3,20.8) 
\DashLine(29.0,20.8)(41.7,20.8){3.0} 
\Text(54.8,11.4)[l]{$W+$}
\DashArrowLine(41.7,20.8)(54.3,11.4){3.0} 
\Text(35,0)[b] {diagr.1}
\end{picture} \ 
%  diagram # 2
\begin{picture}(70,60)(0,0)
\Text(15.9,39.8)[r]{$u$}
\ArrowLine(16.3,39.8)(29.0,39.8) 
\Text(35.1,40.4)[b]{$G$}
\DashLine(29.0,39.8)(41.7,39.8){3.0} 
\Text(54.8,49.3)[l]{$b$}
\ArrowLine(41.7,39.8)(54.3,49.3) 
\Text(54.8,30.3)[l]{$B$}
\ArrowLine(54.3,30.3)(41.7,39.8) 
\Text(27.4,30.3)[r]{$u$}
\ArrowLine(29.0,39.8)(29.0,20.8) 
\Text(15.9,20.8)[r]{$D$}
\ArrowLine(29.0,20.8)(16.3,20.8) 
\DashLine(29.0,20.8)(41.7,20.8){3.0} 
\Text(54.8,11.4)[l]{$W+$}
\DashArrowLine(41.7,20.8)(54.3,11.4){3.0} 
\Text(35,0)[b] {diagr.2}
\end{picture} \ 
%  diagram # 3
\begin{picture}(70,60)(0,0)
\Text(15.9,49.3)[r]{$u$}
\ArrowLine(16.3,49.3)(29.0,39.8) 
\Text(15.9,30.3)[r]{$D$}
\ArrowLine(29.0,39.8)(16.3,30.3) 
\Text(35.1,42.3)[b]{$W+$}
\DashArrowLine(29.0,39.8)(41.7,39.8){3.0} 
\Text(54.8,49.3)[l]{$B$}
\ArrowLine(54.3,49.3)(41.7,39.8) 
\Text(40.0,30.3)[r]{$u$}
\ArrowLine(41.7,39.8)(41.7,20.8) 
\Text(54.8,30.3)[l]{$W+$}
\DashArrowLine(41.7,20.8)(54.3,30.3){3.0} 
\Text(54.8,11.4)[l]{$b$}
\ArrowLine(41.7,20.8)(54.3,11.4) 
\Text(35,0)[b] {diagr.3}
\end{picture} \ 
%  diagram # 4
\begin{picture}(70,60)(0,0)
\Text(15.9,49.3)[r]{$u$}
\ArrowLine(16.3,49.3)(29.0,39.8) 
\Text(15.9,30.3)[r]{$D$}
\ArrowLine(29.0,39.8)(16.3,30.3) 
\Text(35.1,42.3)[b]{$W+$}
\DashArrowLine(29.0,39.8)(41.7,39.8){3.0} 
\Text(54.8,49.3)[l]{$B$}
\ArrowLine(54.3,49.3)(41.7,39.8) 
\Text(40.0,30.3)[r]{$c$}
\ArrowLine(41.7,39.8)(41.7,20.8) 
\Text(54.8,30.3)[l]{$W+$}
\DashArrowLine(41.7,20.8)(54.3,30.3){3.0} 
\Text(54.8,11.4)[l]{$b$}
\ArrowLine(41.7,20.8)(54.3,11.4) 
\Text(35,0)[b] {diagr.4}
\end{picture} \ 
%  diagram # 5
\begin{picture}(70,60)(0,0)
\Text(15.9,49.3)[r]{$u$}
\ArrowLine(16.3,49.3)(29.0,39.8) 
\Text(15.9,30.3)[r]{$D$}
\ArrowLine(29.0,39.8)(16.3,30.3) 
\Text(35.1,42.3)[b]{$W+$}
\DashArrowLine(29.0,39.8)(41.7,39.8){3.0} 
\Text(54.8,49.3)[l]{$B$}
\ArrowLine(54.3,49.3)(41.7,39.8) 
\Text(40.0,30.3)[r]{$t$}
\ArrowLine(41.7,39.8)(41.7,20.8) 
\Text(54.8,30.3)[l]{$W+$}
\DashArrowLine(41.7,20.8)(54.3,30.3){3.0} 
\Text(54.8,11.4)[l]{$b$}
\ArrowLine(41.7,20.8)(54.3,11.4) 
\Text(35,0)[b] {diagr.5}
\end{picture} \ 
%  diagram # 6
\begin{picture}(70,60)(0,0)
\Text(15.9,49.3)[r]{$u$}
\ArrowLine(16.3,49.3)(29.0,39.8) 
\Text(15.9,30.3)[r]{$D$}
\ArrowLine(29.0,39.8)(16.3,30.3) 
\Text(35.1,42.3)[b]{$W+$}
\DashArrowLine(29.0,39.8)(41.7,39.8){3.0} 
\Text(54.8,49.3)[l]{$W+$}
\DashArrowLine(41.7,39.8)(54.3,49.3){3.0} 
\Text(41.3,30.3)[r]{$A$}
\DashLine(41.7,39.8)(41.7,20.8){3.0} 
\Text(54.8,30.3)[l]{$b$}
\ArrowLine(41.7,20.8)(54.3,30.3) 
\Text(54.8,11.4)[l]{$B$}
\ArrowLine(54.3,11.4)(41.7,20.8) 
\Text(35,0)[b] {diagr.6}
\end{picture} \ 
%  diagram # 7
\begin{picture}(70,60)(0,0)
\Text(15.9,49.3)[r]{$u$}
\ArrowLine(16.3,49.3)(29.0,39.8) 
\Text(15.9,30.3)[r]{$D$}
\ArrowLine(29.0,39.8)(16.3,30.3) 
\Text(35.1,42.3)[b]{$W+$}
\DashArrowLine(29.0,39.8)(41.7,39.8){3.0} 
\Text(54.8,49.3)[l]{$W+$}
\DashArrowLine(41.7,39.8)(54.3,49.3){3.0} 
\Text(41.3,30.3)[r]{$H$}
\DashLine(41.7,39.8)(41.7,20.8){1.0}
\Text(54.8,30.3)[l]{$b$}
\ArrowLine(41.7,20.8)(54.3,30.3) 
\Text(54.8,11.4)[l]{$B$}
\ArrowLine(54.3,11.4)(41.7,20.8) 
\Text(35,0)[b] {diagr.7}
\end{picture} \ 
%  diagram # 8
\begin{picture}(70,60)(0,0)
\Text(15.9,49.3)[r]{$u$}
\ArrowLine(16.3,49.3)(29.0,39.8) 
\Text(15.9,30.3)[r]{$D$}
\ArrowLine(29.0,39.8)(16.3,30.3) 
\Text(35.1,42.3)[b]{$W+$}
\DashArrowLine(29.0,39.8)(41.7,39.8){3.0} 
\Text(54.8,49.3)[l]{$W+$}
\DashArrowLine(41.7,39.8)(54.3,49.3){3.0} 
\Text(41.3,30.3)[r]{$Z$}
\DashLine(41.7,39.8)(41.7,20.8){3.0} 
\Text(54.8,30.3)[l]{$b$}
\ArrowLine(41.7,20.8)(54.3,30.3) 
\Text(54.8,11.4)[l]{$B$}
\ArrowLine(54.3,11.4)(41.7,20.8) 
\Text(35,0)[b] {diagr.8}
\end{picture} \ 
%  diagram # 9
\begin{picture}(70,60)(0,0)
\Text(15.9,39.8)[r]{$u$}
\ArrowLine(16.3,39.8)(29.0,39.8) 
\DashLine(29.0,39.8)(41.7,39.8){3.0} 
\Text(54.8,49.3)[l]{$W+$}
\DashArrowLine(41.7,39.8)(54.3,49.3){3.0} 
\Text(27.4,30.3)[r]{$d$}
\ArrowLine(29.0,39.8)(29.0,20.8) 
\Text(15.9,20.8)[r]{$D$}
\ArrowLine(29.0,20.8)(16.3,20.8) 
\Text(35.1,21.5)[b]{$A$}
\DashLine(29.0,20.8)(41.7,20.8){3.0} 
\Text(54.8,30.3)[l]{$b$}
\ArrowLine(41.7,20.8)(54.3,30.3) 
\Text(54.8,11.4)[l]{$B$}
\ArrowLine(54.3,11.4)(41.7,20.8) 
\Text(35,0)[b] {diagr.9}
\end{picture} \ 
%  diagram # 10
\begin{picture}(70,60)(0,0)
\Text(15.9,39.8)[r]{$u$}
\ArrowLine(16.3,39.8)(29.0,39.8) 
\DashLine(29.0,39.8)(41.7,39.8){3.0} 
\Text(54.8,49.3)[l]{$W+$}
\DashArrowLine(41.7,39.8)(54.3,49.3){3.0} 
\Text(27.4,30.3)[r]{$d$}
\ArrowLine(29.0,39.8)(29.0,20.8) 
\Text(15.9,20.8)[r]{$D$}
\ArrowLine(29.0,20.8)(16.3,20.8) 
\Text(35.1,21.5)[b]{$G$}
\DashLine(29.0,20.8)(41.7,20.8){3.0} 
\Text(54.8,30.3)[l]{$b$}
\ArrowLine(41.7,20.8)(54.3,30.3) 
\Text(54.8,11.4)[l]{$B$}
\ArrowLine(54.3,11.4)(41.7,20.8) 
\Text(35,0)[b] {diagr.10}
\end{picture} \ 
%  diagram # 11
\begin{picture}(70,60)(0,0)
\Text(15.9,39.8)[r]{$u$}
\ArrowLine(16.3,39.8)(29.0,39.8) 
\DashLine(29.0,39.8)(41.7,39.8){3.0} 
\Text(54.8,49.3)[l]{$W+$}
\DashArrowLine(41.7,39.8)(54.3,49.3){3.0} 
\Text(27.4,30.3)[r]{$d$}
\ArrowLine(29.0,39.8)(29.0,20.8) 
\Text(15.9,20.8)[r]{$D$}
\ArrowLine(29.0,20.8)(16.3,20.8) 
\Text(35.1,21.5)[b]{$Z$}
\DashLine(29.0,20.8)(41.7,20.8){3.0} 
\Text(54.8,30.3)[l]{$b$}
\ArrowLine(41.7,20.8)(54.3,30.3) 
\Text(54.8,11.4)[l]{$B$}
\ArrowLine(54.3,11.4)(41.7,20.8) 
\Text(35,0)[b] {diagr.11}
\end{picture} \ 
%  diagram # 12
\begin{picture}(70,60)(0,0)
\Text(15.9,49.3)[r]{$u$}
\ArrowLine(16.3,49.3)(41.7,49.3) 
\Text(54.8,49.3)[l]{$b$}
\ArrowLine(41.7,49.3)(54.3,49.3) 
\Text(40.0,39.8)[r]{$W+$}
\DashArrowLine(41.7,49.3)(41.7,30.3){3.0} 
\Text(54.8,30.3)[l]{$B$}
\ArrowLine(54.3,30.3)(41.7,30.3) 
\Text(40.0,20.8)[r]{$u$}
\ArrowLine(41.7,30.3)(41.7,11.4) 
\Text(15.9,11.4)[r]{$D$}
\ArrowLine(41.7,11.4)(16.3,11.4) 
\Text(54.8,11.4)[l]{$W+$}
\DashArrowLine(41.7,11.4)(54.3,11.4){3.0} 
\Text(35,0)[b] {diagr.12}
\end{picture} \ 
%  diagram # 13
\begin{picture}(70,60)(0,0)
\Text(15.9,49.3)[r]{$u$}
\ArrowLine(16.3,49.3)(41.7,49.3) 
\Text(54.8,49.3)[l]{$b$}
\ArrowLine(41.7,49.3)(54.3,49.3) 
\Text(40.0,39.8)[r]{$W+$}
\DashArrowLine(41.7,49.3)(41.7,30.3){3.0} 
\Text(54.8,30.3)[l]{$B$}
\ArrowLine(54.3,30.3)(41.7,30.3) 
\Text(40.0,20.8)[r]{$c$}
\ArrowLine(41.7,30.3)(41.7,11.4) 
\Text(15.9,11.4)[r]{$D$}
\ArrowLine(41.7,11.4)(16.3,11.4) 
\Text(54.8,11.4)[l]{$W+$}
\DashArrowLine(41.7,11.4)(54.3,11.4){3.0} 
\Text(35,0)[b] {diagr.13}
\end{picture} \ 
%  diagram # 14
\begin{picture}(70,60)(0,0)
\Text(15.9,49.3)[r]{$u$}
\ArrowLine(16.3,49.3)(41.7,49.3) 
\Text(54.8,49.3)[l]{$b$}
\ArrowLine(41.7,49.3)(54.3,49.3) 
\Text(40.0,39.8)[r]{$W+$}
\DashArrowLine(41.7,49.3)(41.7,30.3){3.0} 
\Text(54.8,30.3)[l]{$B$}
\ArrowLine(54.3,30.3)(41.7,30.3) 
\Text(40.0,20.8)[r]{$t$}
\ArrowLine(41.7,30.3)(41.7,11.4) 
\Text(15.9,11.4)[r]{$D$}
\ArrowLine(41.7,11.4)(16.3,11.4) 
\Text(54.8,11.4)[l]{$W+$}
\DashArrowLine(41.7,11.4)(54.3,11.4){3.0} 
\Text(35,0)[b] {diagr.14}
\end{picture} \ 
%  diagram # 15
\begin{picture}(70,60)(0,0)
\Text(15.9,39.8)[r]{$u$}
\ArrowLine(16.3,39.8)(29.0,39.8) 
\Text(35.1,40.4)[b]{$Z$}
\DashLine(29.0,39.8)(41.7,39.8){3.0} 
\Text(54.8,49.3)[l]{$b$}
\ArrowLine(41.7,39.8)(54.3,49.3) 
\Text(54.8,30.3)[l]{$B$}
\ArrowLine(54.3,30.3)(41.7,39.8) 
\Text(27.4,30.3)[r]{$u$}
\ArrowLine(29.0,39.8)(29.0,20.8) 
\Text(15.9,20.8)[r]{$D$}
\ArrowLine(29.0,20.8)(16.3,20.8) 
\DashLine(29.0,20.8)(41.7,20.8){3.0} 
\Text(54.8,11.4)[l]{$W+$}
\DashArrowLine(41.7,20.8)(54.3,11.4){3.0} 
\Text(35,0)[b] {diagr.15}
\end{picture} \ 
}
  \end{center}
  \caption{\label{fig:feyn-bbw}%
    Feynman diagrams for the process $u\bar{d}\to b\bar{b}W$.}
\end{figure}
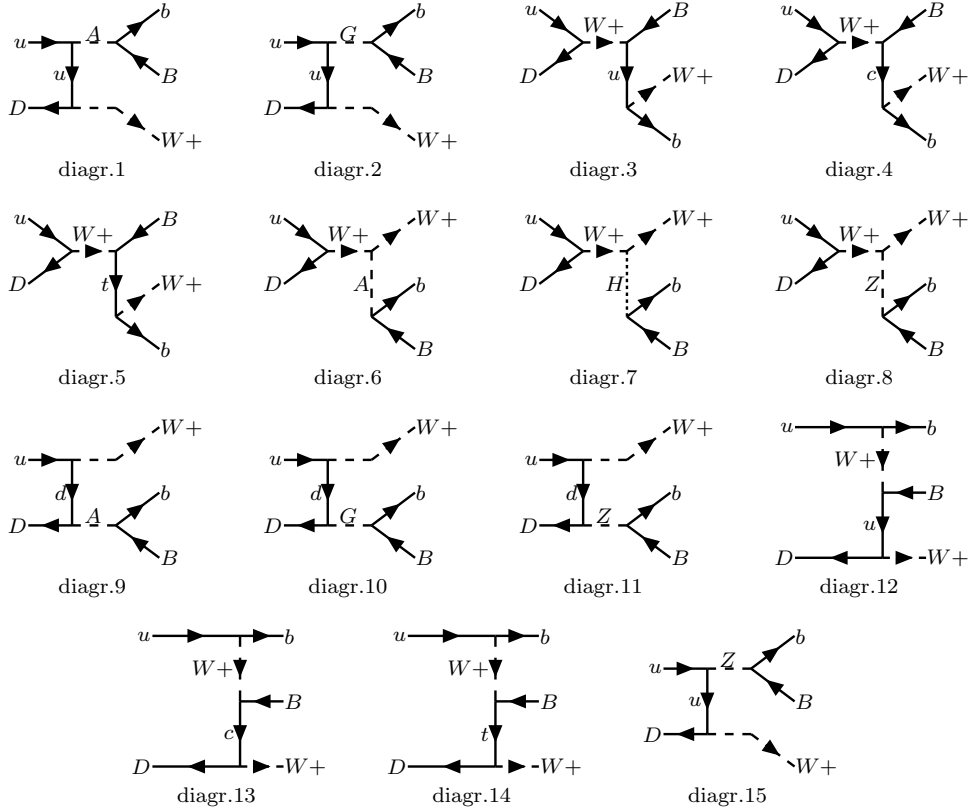
\begin{figure}
  \begin{center}
% CompHEP  version  3.2    
% diagrams for process   PROCESS:  u,D -> G,b,B,W+               74 diagrams
{\def\chepscale{0.89} % picture size control
\def\chepscale{0.7} % picture size control
\unitlength=\chepscale pt
\SetWidth{0.7}      % line    size control
\SetScale{\chepscale}
\tiny   %  letter  size control
%  diagram # 1
\begin{picture}(70,80)(0,0)
\Text(15.9,59.5)[r]{$u$}
\ArrowLine(16.3,59.5)(29.0,59.5) 
\Text(35.1,60.2)[b]{$A$}
\DashLine(29.0,59.5)(41.7,59.5){3.0} 
\Text(54.8,69.1)[l]{$b$}
\ArrowLine(41.7,59.5)(54.3,69.1) 
\Text(54.8,49.9)[l]{$B$}
\ArrowLine(54.3,49.9)(41.7,59.5) 
\Text(27.4,49.9)[r]{$u$}
\ArrowLine(29.0,59.5)(29.0,40.3) 
\DashLine(29.0,40.3)(41.7,40.3){3.0} 
\Text(54.8,30.7)[l]{$G$}
\DashLine(41.7,40.3)(54.3,30.7){3.0} 
\Text(27.4,30.7)[r]{$u$}
\ArrowLine(29.0,40.3)(29.0,21.1) 
\Text(15.9,21.1)[r]{$D$}
\ArrowLine(29.0,21.1)(16.3,21.1) 
\DashLine(29.0,21.1)(41.7,21.1){3.0} 
\Text(54.8,11.5)[l]{$W+$}
\DashArrowLine(41.7,21.1)(54.3,11.5){3.0} 
\Text(35,0)[b] {diagr.1}
\end{picture} \ 
%  diagram # 2
\begin{picture}(70,80)(0,0)
\Text(15.9,59.5)[r]{$u$}
\ArrowLine(16.3,59.5)(29.0,59.5) 
\Text(35.1,60.2)[b]{$A$}
\DashLine(29.0,59.5)(41.7,59.5){3.0} 
\Text(54.8,69.1)[l]{$b$}
\ArrowLine(41.7,59.5)(54.3,69.1) 
\Text(54.8,49.9)[l]{$B$}
\ArrowLine(54.3,49.9)(41.7,59.5) 
\Text(27.4,49.9)[r]{$u$}
\ArrowLine(29.0,59.5)(29.0,40.3) 
\DashLine(29.0,40.3)(41.7,40.3){3.0} 
\Text(54.8,30.7)[l]{$W+$}
\DashArrowLine(41.7,40.3)(54.3,30.7){3.0} 
\Text(27.4,30.7)[r]{$d$}
\ArrowLine(29.0,40.3)(29.0,21.1) 
\Text(15.9,21.1)[r]{$D$}
\ArrowLine(29.0,21.1)(16.3,21.1) 
\DashLine(29.0,21.1)(41.7,21.1){3.0} 
\Text(54.8,11.5)[l]{$G$}
\DashLine(41.7,21.1)(54.3,11.5){3.0} 
\Text(35,0)[b] {diagr.2}
\end{picture} \ 
%  diagram # 3
\begin{picture}(70,80)(0,0)
\Text(15.9,59.5)[r]{$u$}
\ArrowLine(16.3,59.5)(29.0,59.5) 
\Text(35.1,60.2)[b]{$A$}
\DashLine(29.0,59.5)(41.7,59.5){3.0} 
\Text(54.8,69.1)[l]{$B$}
\ArrowLine(54.3,69.1)(41.7,59.5) 
\Text(40.0,49.9)[r]{$b$}
\ArrowLine(41.7,59.5)(41.7,40.3) 
\Text(54.8,49.9)[l]{$b$}
\ArrowLine(41.7,40.3)(54.3,49.9) 
\Text(54.8,30.7)[l]{$G$}
\DashLine(41.7,40.3)(54.3,30.7){3.0} 
\Text(27.4,40.3)[r]{$u$}
\ArrowLine(29.0,59.5)(29.0,21.1) 
\Text(15.9,21.1)[r]{$D$}
\ArrowLine(29.0,21.1)(16.3,21.1) 
\DashLine(29.0,21.1)(41.7,21.1){3.0} 
\Text(54.8,11.5)[l]{$W+$}
\DashArrowLine(41.7,21.1)(54.3,11.5){3.0} 
\Text(35,0)[b] {diagr.3}
\end{picture} \ 
%  diagram # 4
\begin{picture}(70,80)(0,0)
\Text(15.9,59.5)[r]{$u$}
\ArrowLine(16.3,59.5)(29.0,59.5) 
\Text(35.1,60.2)[b]{$A$}
\DashLine(29.0,59.5)(41.7,59.5){3.0} 
\Text(54.8,69.1)[l]{$b$}
\ArrowLine(41.7,59.5)(54.3,69.1) 
\Text(40.0,49.9)[r]{$b$}
\ArrowLine(41.7,40.3)(41.7,59.5) 
\Text(54.8,49.9)[l]{$B$}
\ArrowLine(54.3,49.9)(41.7,40.3) 
\Text(54.8,30.7)[l]{$G$}
\DashLine(41.7,40.3)(54.3,30.7){3.0} 
\Text(27.4,40.3)[r]{$u$}
\ArrowLine(29.0,59.5)(29.0,21.1) 
\Text(15.9,21.1)[r]{$D$}
\ArrowLine(29.0,21.1)(16.3,21.1) 
\DashLine(29.0,21.1)(41.7,21.1){3.0} 
\Text(54.8,11.5)[l]{$W+$}
\DashArrowLine(41.7,21.1)(54.3,11.5){3.0} 
\Text(35,0)[b] {diagr.4}
\end{picture} \ 
%  diagram # 5
\begin{picture}(70,80)(0,0)
\Text(15.9,59.5)[r]{$u$}
\ArrowLine(16.3,59.5)(29.0,59.5) 
\DashLine(29.0,59.5)(41.7,59.5){3.0} 
\Text(54.8,69.1)[l]{$G$}
\DashLine(41.7,59.5)(54.3,69.1){3.0} 
\Text(27.4,49.9)[r]{$u$}
\ArrowLine(29.0,59.5)(29.0,40.3) 
\Text(35.1,41.0)[b]{$A$}
\DashLine(29.0,40.3)(41.7,40.3){3.0} 
\Text(54.8,49.9)[l]{$b$}
\ArrowLine(41.7,40.3)(54.3,49.9) 
\Text(54.8,30.7)[l]{$B$}
\ArrowLine(54.3,30.7)(41.7,40.3) 
\Text(27.4,30.7)[r]{$u$}
\ArrowLine(29.0,40.3)(29.0,21.1) 
\Text(15.9,21.1)[r]{$D$}
\ArrowLine(29.0,21.1)(16.3,21.1) 
\DashLine(29.0,21.1)(41.7,21.1){3.0} 
\Text(54.8,11.5)[l]{$W+$}
\DashArrowLine(41.7,21.1)(54.3,11.5){3.0} 
\Text(35,0)[b] {diagr.5}
\end{picture} \ 
%  diagram # 6
\begin{picture}(70,80)(0,0)
\Text(15.9,59.5)[r]{$u$}
\ArrowLine(16.3,59.5)(29.0,59.5) 
\DashLine(29.0,59.5)(41.7,59.5){3.0} 
\Text(54.8,69.1)[l]{$G$}
\DashLine(41.7,59.5)(54.3,69.1){3.0} 
\Text(27.4,49.9)[r]{$u$}
\ArrowLine(29.0,59.5)(29.0,40.3) 
\Text(35.1,41.0)[b]{$G$}
\DashLine(29.0,40.3)(41.7,40.3){3.0} 
\Text(54.8,49.9)[l]{$b$}
\ArrowLine(41.7,40.3)(54.3,49.9) 
\Text(54.8,30.7)[l]{$B$}
\ArrowLine(54.3,30.7)(41.7,40.3) 
\Text(27.4,30.7)[r]{$u$}
\ArrowLine(29.0,40.3)(29.0,21.1) 
\Text(15.9,21.1)[r]{$D$}
\ArrowLine(29.0,21.1)(16.3,21.1) 
\DashLine(29.0,21.1)(41.7,21.1){3.0} 
\Text(54.8,11.5)[l]{$W+$}
\DashArrowLine(41.7,21.1)(54.3,11.5){3.0} 
\Text(35,0)[b] {diagr.6}
\end{picture} \ 
%  diagram # 9
\begin{picture}(70,80)(0,0)
\Text(15.9,59.5)[r]{$u$}
\ArrowLine(16.3,59.5)(29.0,59.5) 
\DashLine(29.0,59.5)(41.7,59.5){3.0} 
\Text(54.8,69.1)[l]{$G$}
\DashLine(41.7,59.5)(54.3,69.1){3.0} 
\Text(27.4,49.9)[r]{$u$}
\ArrowLine(29.0,59.5)(29.0,40.3) 
\Text(15.9,40.3)[r]{$D$}
\ArrowLine(29.0,40.3)(16.3,40.3) 
\Text(35.1,42.9)[b]{$W+$}
\DashArrowLine(29.0,40.3)(41.7,40.3){3.0} 
\Text(54.8,49.9)[l]{$B$}
\ArrowLine(54.3,49.9)(41.7,40.3) 
\Text(40.0,30.7)[r]{$t$}
\ArrowLine(41.7,40.3)(41.7,21.1) 
\Text(54.8,30.7)[l]{$W+$}
\DashArrowLine(41.7,21.1)(54.3,30.7){3.0} 
\Text(54.8,11.5)[l]{$b$}
\ArrowLine(41.7,21.1)(54.3,11.5) 
\Text(35,0)[b] {diagr.7}
\end{picture} \ 
%  diagram # 10
\begin{picture}(70,80)(0,0)
\Text(15.9,59.5)[r]{$u$}
\ArrowLine(16.3,59.5)(29.0,59.5) 
\DashLine(29.0,59.5)(41.7,59.5){3.0} 
\Text(54.8,69.1)[l]{$G$}
\DashLine(41.7,59.5)(54.3,69.1){3.0} 
\Text(27.4,49.9)[r]{$u$}
\ArrowLine(29.0,59.5)(29.0,40.3) 
\Text(15.9,40.3)[r]{$D$}
\ArrowLine(29.0,40.3)(16.3,40.3) 
\Text(35.1,42.9)[b]{$W+$}
\DashArrowLine(29.0,40.3)(41.7,40.3){3.0} 
\Text(54.8,49.9)[l]{$W+$}
\DashArrowLine(41.7,40.3)(54.3,49.9){3.0} 
\Text(41.3,30.7)[r]{$A$}
\DashLine(41.7,40.3)(41.7,21.1){3.0} 
\Text(54.8,30.7)[l]{$b$}
\ArrowLine(41.7,21.1)(54.3,30.7) 
\Text(54.8,11.5)[l]{$B$}
\ArrowLine(54.3,11.5)(41.7,21.1) 
\Text(35,0)[b] {diagr.8}
\end{picture} \ 
%  diagram # 11
\begin{picture}(70,80)(0,0)
\Text(15.9,59.5)[r]{$u$}
\ArrowLine(16.3,59.5)(29.0,59.5) 
\DashLine(29.0,59.5)(41.7,59.5){3.0} 
\Text(54.8,69.1)[l]{$G$}
\DashLine(41.7,59.5)(54.3,69.1){3.0} 
\Text(27.4,49.9)[r]{$u$}
\ArrowLine(29.0,59.5)(29.0,40.3) 
\Text(15.9,40.3)[r]{$D$}
\ArrowLine(29.0,40.3)(16.3,40.3) 
\Text(35.1,42.9)[b]{$W+$}
\DashArrowLine(29.0,40.3)(41.7,40.3){3.0} 
\Text(54.8,49.9)[l]{$W+$}
\DashArrowLine(41.7,40.3)(54.3,49.9){3.0} 
\Text(41.3,30.7)[r]{$H$}
\DashLine(41.7,40.3)(41.7,21.1){1.0}
\Text(54.8,30.7)[l]{$b$}
\ArrowLine(41.7,21.1)(54.3,30.7) 
\Text(54.8,11.5)[l]{$B$}
\ArrowLine(54.3,11.5)(41.7,21.1) 
\Text(35,0)[b] {diagr.9}
\end{picture} \ 
%  diagram # 12
\begin{picture}(70,80)(0,0)
\Text(15.9,59.5)[r]{$u$}
\ArrowLine(16.3,59.5)(29.0,59.5) 
\DashLine(29.0,59.5)(41.7,59.5){3.0} 
\Text(54.8,69.1)[l]{$G$}
\DashLine(41.7,59.5)(54.3,69.1){3.0} 
\Text(27.4,49.9)[r]{$u$}
\ArrowLine(29.0,59.5)(29.0,40.3) 
\Text(15.9,40.3)[r]{$D$}
\ArrowLine(29.0,40.3)(16.3,40.3) 
\Text(35.1,42.9)[b]{$W+$}
\DashArrowLine(29.0,40.3)(41.7,40.3){3.0} 
\Text(54.8,49.9)[l]{$W+$}
\DashArrowLine(41.7,40.3)(54.3,49.9){3.0} 
\Text(41.3,30.7)[r]{$Z$}
\DashLine(41.7,40.3)(41.7,21.1){3.0} 
\Text(54.8,30.7)[l]{$b$}
\ArrowLine(41.7,21.1)(54.3,30.7) 
\Text(54.8,11.5)[l]{$B$}
\ArrowLine(54.3,11.5)(41.7,21.1) 
\Text(35,0)[b] {diagr.10}
\end{picture} \ 
%  diagram # 13
\begin{picture}(70,80)(0,0)
\Text(15.9,59.5)[r]{$u$}
\ArrowLine(16.3,59.5)(29.0,59.5) 
\DashLine(29.0,59.5)(41.7,59.5){3.0} 
\Text(54.8,69.1)[l]{$G$}
\DashLine(41.7,59.5)(54.3,69.1){3.0} 
\Text(27.4,49.9)[r]{$u$}
\ArrowLine(29.0,59.5)(29.0,40.3) 
\DashLine(29.0,40.3)(41.7,40.3){3.0} 
\Text(54.8,49.9)[l]{$W+$}
\DashArrowLine(41.7,40.3)(54.3,49.9){3.0} 
\Text(27.4,30.7)[r]{$d$}
\ArrowLine(29.0,40.3)(29.0,21.1) 
\Text(15.9,21.1)[r]{$D$}
\ArrowLine(29.0,21.1)(16.3,21.1) 
\Text(35.1,21.8)[b]{$A$}
\DashLine(29.0,21.1)(41.7,21.1){3.0} 
\Text(54.8,30.7)[l]{$b$}
\ArrowLine(41.7,21.1)(54.3,30.7) 
\Text(54.8,11.5)[l]{$B$}
\ArrowLine(54.3,11.5)(41.7,21.1) 
\Text(35,0)[b] {diagr.11}
\end{picture} \ 
%  diagram # 14
\begin{picture}(70,80)(0,0)
\Text(15.9,59.5)[r]{$u$}
\ArrowLine(16.3,59.5)(29.0,59.5) 
\DashLine(29.0,59.5)(41.7,59.5){3.0} 
\Text(54.8,69.1)[l]{$G$}
\DashLine(41.7,59.5)(54.3,69.1){3.0} 
\Text(27.4,49.9)[r]{$u$}
\ArrowLine(29.0,59.5)(29.0,40.3) 
\DashLine(29.0,40.3)(41.7,40.3){3.0} 
\Text(54.8,49.9)[l]{$W+$}
\DashArrowLine(41.7,40.3)(54.3,49.9){3.0} 
\Text(27.4,30.7)[r]{$d$}
\ArrowLine(29.0,40.3)(29.0,21.1) 
\Text(15.9,21.1)[r]{$D$}
\ArrowLine(29.0,21.1)(16.3,21.1) 
\Text(35.1,21.8)[b]{$G$}
\DashLine(29.0,21.1)(41.7,21.1){3.0} 
\Text(54.8,30.7)[l]{$b$}
\ArrowLine(41.7,21.1)(54.3,30.7) 
\Text(54.8,11.5)[l]{$B$}
\ArrowLine(54.3,11.5)(41.7,21.1) 
\Text(35,0)[b] {diagr.12}
\end{picture} \ 
%  diagram # 15
\begin{picture}(70,80)(0,0)
\Text(15.9,59.5)[r]{$u$}
\ArrowLine(16.3,59.5)(29.0,59.5) 
\DashLine(29.0,59.5)(41.7,59.5){3.0} 
\Text(54.8,69.1)[l]{$G$}
\DashLine(41.7,59.5)(54.3,69.1){3.0} 
\Text(27.4,49.9)[r]{$u$}
\ArrowLine(29.0,59.5)(29.0,40.3) 
\DashLine(29.0,40.3)(41.7,40.3){3.0} 
\Text(54.8,49.9)[l]{$W+$}
\DashArrowLine(41.7,40.3)(54.3,49.9){3.0} 
\Text(27.4,30.7)[r]{$d$}
\ArrowLine(29.0,40.3)(29.0,21.1) 
\Text(15.9,21.1)[r]{$D$}
\ArrowLine(29.0,21.1)(16.3,21.1) 
\Text(35.1,21.8)[b]{$Z$}
\DashLine(29.0,21.1)(41.7,21.1){3.0} 
\Text(54.8,30.7)[l]{$b$}
\ArrowLine(41.7,21.1)(54.3,30.7) 
\Text(54.8,11.5)[l]{$B$}
\ArrowLine(54.3,11.5)(41.7,21.1) 
\Text(35,0)[b] {diagr.13}
\end{picture} \ 
%  diagram # 18
\begin{picture}(70,80)(0,0)
\Text(15.9,69.1)[r]{$u$}
\ArrowLine(16.3,69.1)(41.7,69.1) 
\Text(54.8,69.1)[l]{$G$}
\DashLine(41.7,69.1)(54.3,69.1){3.0} 
\Text(40.0,59.5)[r]{$u$}
\ArrowLine(41.7,69.1)(41.7,49.9) 
\Text(54.8,49.9)[l]{$b$}
\ArrowLine(41.7,49.9)(54.3,49.9) 
\Text(40.0,40.3)[r]{$W+$}
\DashArrowLine(41.7,49.9)(41.7,30.7){3.0} 
\Text(54.8,30.7)[l]{$B$}
\ArrowLine(54.3,30.7)(41.7,30.7) 
\Text(40.0,21.1)[r]{$t$}
\ArrowLine(41.7,30.7)(41.7,11.5) 
\Text(15.9,11.5)[r]{$D$}
\ArrowLine(41.7,11.5)(16.3,11.5) 
\Text(54.8,11.5)[l]{$W+$}
\DashArrowLine(41.7,11.5)(54.3,11.5){3.0} 
\Text(35,0)[b] {diagr.14}
\end{picture} \ 
%  diagram # 19
\begin{picture}(70,80)(0,0)
\Text(15.9,59.5)[r]{$u$}
\ArrowLine(16.3,59.5)(29.0,59.5) 
\DashLine(29.0,59.5)(41.7,59.5){3.0} 
\Text(54.8,69.1)[l]{$G$}
\DashLine(41.7,59.5)(54.3,69.1){3.0} 
\Text(27.4,49.9)[r]{$u$}
\ArrowLine(29.0,59.5)(29.0,40.3) 
\Text(35.1,41.0)[b]{$Z$}
\DashLine(29.0,40.3)(41.7,40.3){3.0} 
\Text(54.8,49.9)[l]{$b$}
\ArrowLine(41.7,40.3)(54.3,49.9) 
\Text(54.8,30.7)[l]{$B$}
\ArrowLine(54.3,30.7)(41.7,40.3) 
\Text(27.4,30.7)[r]{$u$}
\ArrowLine(29.0,40.3)(29.0,21.1) 
\Text(15.9,21.1)[r]{$D$}
\ArrowLine(29.0,21.1)(16.3,21.1) 
\DashLine(29.0,21.1)(41.7,21.1){3.0} 
\Text(54.8,11.5)[l]{$W+$}
\DashArrowLine(41.7,21.1)(54.3,11.5){3.0} 
\Text(35,0)[b] {diagr.15}
\end{picture} \ 
%  diagram # 20
\begin{picture}(70,80)(0,0)
\Text(15.9,59.5)[r]{$u$}
\ArrowLine(16.3,59.5)(29.0,59.5) 
\Text(35.1,60.2)[b]{$G$}
\DashLine(29.0,59.5)(41.7,59.5){3.0} 
\Text(54.8,69.1)[l]{$b$}
\ArrowLine(41.7,59.5)(54.3,69.1) 
\Text(54.8,49.9)[l]{$B$}
\ArrowLine(54.3,49.9)(41.7,59.5) 
\Text(27.4,49.9)[r]{$u$}
\ArrowLine(29.0,59.5)(29.0,40.3) 
\DashLine(29.0,40.3)(41.7,40.3){3.0} 
\Text(54.8,30.7)[l]{$G$}
\DashLine(41.7,40.3)(54.3,30.7){3.0} 
\Text(27.4,30.7)[r]{$u$}
\ArrowLine(29.0,40.3)(29.0,21.1) 
\Text(15.9,21.1)[r]{$D$}
\ArrowLine(29.0,21.1)(16.3,21.1) 
\DashLine(29.0,21.1)(41.7,21.1){3.0} 
\Text(54.8,11.5)[l]{$W+$}
\DashArrowLine(41.7,21.1)(54.3,11.5){3.0} 
\Text(35,0)[b] {diagr.16}
\end{picture} \ 
%  diagram # 21
\begin{picture}(70,80)(0,0)
\Text(15.9,59.5)[r]{$u$}
\ArrowLine(16.3,59.5)(29.0,59.5) 
\Text(35.1,60.2)[b]{$G$}
\DashLine(29.0,59.5)(41.7,59.5){3.0} 
\Text(54.8,69.1)[l]{$b$}
\ArrowLine(41.7,59.5)(54.3,69.1) 
\Text(54.8,49.9)[l]{$B$}
\ArrowLine(54.3,49.9)(41.7,59.5) 
\Text(27.4,49.9)[r]{$u$}
\ArrowLine(29.0,59.5)(29.0,40.3) 
\DashLine(29.0,40.3)(41.7,40.3){3.0} 
\Text(54.8,30.7)[l]{$W+$}
\DashArrowLine(41.7,40.3)(54.3,30.7){3.0} 
\Text(27.4,30.7)[r]{$d$}
\ArrowLine(29.0,40.3)(29.0,21.1) 
\Text(15.9,21.1)[r]{$D$}
\ArrowLine(29.0,21.1)(16.3,21.1) 
\DashLine(29.0,21.1)(41.7,21.1){3.0} 
\Text(54.8,11.5)[l]{$G$}
\DashLine(41.7,21.1)(54.3,11.5){3.0} 
\Text(35,0)[b] {diagr.17}
\end{picture} \ 
%  diagram # 22
\begin{picture}(70,80)(0,0)
\Text(15.9,59.5)[r]{$u$}
\ArrowLine(16.3,59.5)(29.0,59.5) 
\Text(35.1,60.2)[b]{$G$}
\DashLine(29.0,59.5)(41.7,59.5){3.0} 
\Text(54.8,69.1)[l]{$G$}
\DashLine(41.7,59.5)(54.3,69.1){3.0} 
\Text(41.3,49.9)[r]{$G$}
\DashLine(41.7,59.5)(41.7,40.3){3.0} 
\Text(54.8,49.9)[l]{$b$}
\ArrowLine(41.7,40.3)(54.3,49.9) 
\Text(54.8,30.7)[l]{$B$}
\ArrowLine(54.3,30.7)(41.7,40.3) 
\Text(27.4,40.3)[r]{$u$}
\ArrowLine(29.0,59.5)(29.0,21.1) 
\Text(15.9,21.1)[r]{$D$}
\ArrowLine(29.0,21.1)(16.3,21.1) 
\DashLine(29.0,21.1)(41.7,21.1){3.0} 
\Text(54.8,11.5)[l]{$W+$}
\DashArrowLine(41.7,21.1)(54.3,11.5){3.0} 
\Text(35,0)[b] {diagr.18}
\end{picture} \ 
%  diagram # 23
\begin{picture}(70,80)(0,0)
\Text(15.9,59.5)[r]{$u$}
\ArrowLine(16.3,59.5)(29.0,59.5) 
\Text(35.1,60.2)[b]{$G$}
\DashLine(29.0,59.5)(41.7,59.5){3.0} 
\Text(54.8,69.1)[l]{$B$}
\ArrowLine(54.3,69.1)(41.7,59.5) 
\Text(40.0,49.9)[r]{$b$}
\ArrowLine(41.7,59.5)(41.7,40.3) 
\Text(54.8,49.9)[l]{$b$}
\ArrowLine(41.7,40.3)(54.3,49.9) 
\Text(54.8,30.7)[l]{$G$}
\DashLine(41.7,40.3)(54.3,30.7){3.0} 
\Text(27.4,40.3)[r]{$u$}
\ArrowLine(29.0,59.5)(29.0,21.1) 
\Text(15.9,21.1)[r]{$D$}
\ArrowLine(29.0,21.1)(16.3,21.1) 
\DashLine(29.0,21.1)(41.7,21.1){3.0} 
\Text(54.8,11.5)[l]{$W+$}
\DashArrowLine(41.7,21.1)(54.3,11.5){3.0} 
\Text(35,0)[b] {diagr.19}
\end{picture} \ 
%  diagram # 24
\begin{picture}(70,80)(0,0)
\Text(15.9,59.5)[r]{$u$}
\ArrowLine(16.3,59.5)(29.0,59.5) 
\Text(35.1,60.2)[b]{$G$}
\DashLine(29.0,59.5)(41.7,59.5){3.0} 
\Text(54.8,69.1)[l]{$b$}
\ArrowLine(41.7,59.5)(54.3,69.1) 
\Text(40.0,49.9)[r]{$b$}
\ArrowLine(41.7,40.3)(41.7,59.5) 
\Text(54.8,49.9)[l]{$B$}
\ArrowLine(54.3,49.9)(41.7,40.3) 
\Text(54.8,30.7)[l]{$G$}
\DashLine(41.7,40.3)(54.3,30.7){3.0} 
\Text(27.4,40.3)[r]{$u$}
\ArrowLine(29.0,59.5)(29.0,21.1) 
\Text(15.9,21.1)[r]{$D$}
\ArrowLine(29.0,21.1)(16.3,21.1) 
\DashLine(29.0,21.1)(41.7,21.1){3.0} 
\Text(54.8,11.5)[l]{$W+$}
\DashArrowLine(41.7,21.1)(54.3,11.5){3.0} 
\Text(35,0)[b] {diagr.20}
\end{picture} \ 
%  diagram # 31
\begin{picture}(70,80)(0,0)
\Text(15.9,49.9)[r]{$u$}
\ArrowLine(16.3,49.9)(29.0,40.3) 
\Text(15.9,30.7)[r]{$D$}
\ArrowLine(29.0,40.3)(16.3,30.7) 
\Text(29.0,41.6)[lb]{$W+$}
\DashArrowLine(29.0,40.3)(41.7,40.3){3.0} 
\Text(40.0,49.9)[r]{$t$}
\ArrowLine(41.7,40.3)(41.7,59.5) 
\Text(54.8,69.1)[l]{$W+$}
\DashArrowLine(41.7,59.5)(54.3,69.1){3.0} 
\Text(54.8,49.9)[l]{$b$}
\ArrowLine(41.7,59.5)(54.3,49.9) 
\Text(40.0,30.7)[r]{$b$}
\ArrowLine(41.7,21.1)(41.7,40.3) 
\Text(54.8,30.7)[l]{$B$}
\ArrowLine(54.3,30.7)(41.7,21.1) 
\Text(54.8,11.5)[l]{$G$}
\DashLine(41.7,21.1)(54.3,11.5){3.0} 
\Text(35,0)[b] {diagr.21}
\end{picture} \ 
%  diagram # 32
\begin{picture}(70,80)(0,0)
\Text(15.9,69.1)[r]{$u$}
\ArrowLine(16.3,69.1)(29.0,59.5) 
\Text(15.9,49.9)[r]{$D$}
\ArrowLine(29.0,59.5)(16.3,49.9) 
\Text(35.1,62.1)[b]{$W+$}
\DashArrowLine(29.0,59.5)(41.7,59.5){3.0} 
\Text(54.8,69.1)[l]{$B$}
\ArrowLine(54.3,69.1)(41.7,59.5) 
\Text(40.0,49.9)[r]{$t$}
\ArrowLine(41.7,59.5)(41.7,40.3) 
\Text(54.8,49.9)[l]{$G$}
\DashLine(41.7,40.3)(54.3,49.9){3.0} 
\Text(40.0,30.7)[r]{$t$}
\ArrowLine(41.7,40.3)(41.7,21.1) 
\Text(54.8,30.7)[l]{$W+$}
\DashArrowLine(41.7,21.1)(54.3,30.7){3.0} 
\Text(54.8,11.5)[l]{$b$}
\ArrowLine(41.7,21.1)(54.3,11.5) 
\Text(35,0)[b] {diagr.22}
\end{picture} \ 
%  diagram # 33
\begin{picture}(70,80)(0,0)
\Text(15.9,69.1)[r]{$u$}
\ArrowLine(16.3,69.1)(29.0,59.5) 
\Text(15.9,49.9)[r]{$D$}
\ArrowLine(29.0,59.5)(16.3,49.9) 
\Text(35.1,62.1)[b]{$W+$}
\DashArrowLine(29.0,59.5)(41.7,59.5){3.0} 
\Text(54.8,69.1)[l]{$B$}
\ArrowLine(54.3,69.1)(41.7,59.5) 
\Text(40.0,49.9)[r]{$t$}
\ArrowLine(41.7,59.5)(41.7,40.3) 
\Text(54.8,49.9)[l]{$W+$}
\DashArrowLine(41.7,40.3)(54.3,49.9){3.0} 
\Text(40.0,30.7)[r]{$b$}
\ArrowLine(41.7,40.3)(41.7,21.1) 
\Text(54.8,30.7)[l]{$b$}
\ArrowLine(41.7,21.1)(54.3,30.7) 
\Text(54.8,11.5)[l]{$G$}
\DashLine(41.7,21.1)(54.3,11.5){3.0} 
\Text(35,0)[b] {diagr.23}
\end{picture} \ 
%  diagram # 34
\begin{picture}(70,80)(0,0)
\Text(15.9,69.1)[r]{$u$}
\ArrowLine(16.3,69.1)(29.0,59.5) 
\Text(15.9,49.9)[r]{$D$}
\ArrowLine(29.0,59.5)(16.3,49.9) 
\Text(35.1,62.1)[b]{$W+$}
\DashArrowLine(29.0,59.5)(41.7,59.5){3.0} 
\Text(54.8,69.1)[l]{$W+$}
\DashArrowLine(41.7,59.5)(54.3,69.1){3.0} 
\Text(41.3,49.9)[r]{$A$}
\DashLine(41.7,59.5)(41.7,40.3){3.0} 
\Text(54.8,49.9)[l]{$B$}
\ArrowLine(54.3,49.9)(41.7,40.3) 
\Text(40.0,30.7)[r]{$b$}
\ArrowLine(41.7,40.3)(41.7,21.1) 
\Text(54.8,30.7)[l]{$b$}
\ArrowLine(41.7,21.1)(54.3,30.7) 
\Text(54.8,11.5)[l]{$G$}
\DashLine(41.7,21.1)(54.3,11.5){3.0} 
\Text(35,0)[b] {diagr.24}
\end{picture} \ 
%  diagram # 35
\begin{picture}(70,80)(0,0)
\Text(15.9,69.1)[r]{$u$}
\ArrowLine(16.3,69.1)(29.0,59.5) 
\Text(15.9,49.9)[r]{$D$}
\ArrowLine(29.0,59.5)(16.3,49.9) 
\Text(35.1,62.1)[b]{$W+$}
\DashArrowLine(29.0,59.5)(41.7,59.5){3.0} 
\Text(54.8,69.1)[l]{$W+$}
\DashArrowLine(41.7,59.5)(54.3,69.1){3.0} 
\Text(41.3,49.9)[r]{$A$}
\DashLine(41.7,59.5)(41.7,40.3){3.0} 
\Text(54.8,49.9)[l]{$b$}
\ArrowLine(41.7,40.3)(54.3,49.9) 
\Text(40.0,30.7)[r]{$b$}
\ArrowLine(41.7,21.1)(41.7,40.3) 
\Text(54.8,30.7)[l]{$B$}
\ArrowLine(54.3,30.7)(41.7,21.1) 
\Text(54.8,11.5)[l]{$G$}
\DashLine(41.7,21.1)(54.3,11.5){3.0} 
\Text(35,0)[b] {diagr.25}
\end{picture} \ 
%  diagram # 36
\begin{picture}(70,80)(0,0)
\Text(15.9,69.1)[r]{$u$}
\ArrowLine(16.3,69.1)(29.0,59.5) 
\Text(15.9,49.9)[r]{$D$}
\ArrowLine(29.0,59.5)(16.3,49.9) 
\Text(35.1,62.1)[b]{$W+$}
\DashArrowLine(29.0,59.5)(41.7,59.5){3.0} 
\Text(54.8,69.1)[l]{$W+$}
\DashArrowLine(41.7,59.5)(54.3,69.1){3.0} 
\Text(41.3,49.9)[r]{$H$}
\DashLine(41.7,59.5)(41.7,40.3){1.0}
\Text(54.8,49.9)[l]{$B$}
\ArrowLine(54.3,49.9)(41.7,40.3) 
\Text(40.0,30.7)[r]{$b$}
\ArrowLine(41.7,40.3)(41.7,21.1) 
\Text(54.8,30.7)[l]{$b$}
\ArrowLine(41.7,21.1)(54.3,30.7) 
\Text(54.8,11.5)[l]{$G$}
\DashLine(41.7,21.1)(54.3,11.5){3.0} 
\Text(35,0)[b] {diagr.26}
\end{picture} \ 
%  diagram # 37
\begin{picture}(70,80)(0,0)
\Text(15.9,69.1)[r]{$u$}
\ArrowLine(16.3,69.1)(29.0,59.5) 
\Text(15.9,49.9)[r]{$D$}
\ArrowLine(29.0,59.5)(16.3,49.9) 
\Text(35.1,62.1)[b]{$W+$}
\DashArrowLine(29.0,59.5)(41.7,59.5){3.0} 
\Text(54.8,69.1)[l]{$W+$}
\DashArrowLine(41.7,59.5)(54.3,69.1){3.0} 
\Text(41.3,49.9)[r]{$H$}
\DashLine(41.7,59.5)(41.7,40.3){1.0}
\Text(54.8,49.9)[l]{$b$}
\ArrowLine(41.7,40.3)(54.3,49.9) 
\Text(40.0,30.7)[r]{$b$}
\ArrowLine(41.7,21.1)(41.7,40.3) 
\Text(54.8,30.7)[l]{$B$}
\ArrowLine(54.3,30.7)(41.7,21.1) 
\Text(54.8,11.5)[l]{$G$}
\DashLine(41.7,21.1)(54.3,11.5){3.0} 
\Text(35,0)[b] {diagr.27}
\end{picture} \ 
%  diagram # 38
\begin{picture}(70,80)(0,0)
\Text(15.9,69.1)[r]{$u$}
\ArrowLine(16.3,69.1)(29.0,59.5) 
\Text(15.9,49.9)[r]{$D$}
\ArrowLine(29.0,59.5)(16.3,49.9) 
\Text(35.1,62.1)[b]{$W+$}
\DashArrowLine(29.0,59.5)(41.7,59.5){3.0} 
\Text(54.8,69.1)[l]{$W+$}
\DashArrowLine(41.7,59.5)(54.3,69.1){3.0} 
\Text(41.3,49.9)[r]{$Z$}
\DashLine(41.7,59.5)(41.7,40.3){3.0} 
\Text(54.8,49.9)[l]{$B$}
\ArrowLine(54.3,49.9)(41.7,40.3) 
\Text(40.0,30.7)[r]{$b$}
\ArrowLine(41.7,40.3)(41.7,21.1) 
\Text(54.8,30.7)[l]{$b$}
\ArrowLine(41.7,21.1)(54.3,30.7) 
\Text(54.8,11.5)[l]{$G$}
\DashLine(41.7,21.1)(54.3,11.5){3.0} 
\Text(35,0)[b] {diagr.28}
\end{picture} \ 
%  diagram # 39
\begin{picture}(70,80)(0,0)
\Text(15.9,69.1)[r]{$u$}
\ArrowLine(16.3,69.1)(29.0,59.5) 
\Text(15.9,49.9)[r]{$D$}
\ArrowLine(29.0,59.5)(16.3,49.9) 
\Text(35.1,62.1)[b]{$W+$}
\DashArrowLine(29.0,59.5)(41.7,59.5){3.0} 
\Text(54.8,69.1)[l]{$W+$}
\DashArrowLine(41.7,59.5)(54.3,69.1){3.0} 
\Text(41.3,49.9)[r]{$Z$}
\DashLine(41.7,59.5)(41.7,40.3){3.0} 
\Text(54.8,49.9)[l]{$b$}
\ArrowLine(41.7,40.3)(54.3,49.9) 
\Text(40.0,30.7)[r]{$b$}
\ArrowLine(41.7,21.1)(41.7,40.3) 
\Text(54.8,30.7)[l]{$B$}
\ArrowLine(54.3,30.7)(41.7,21.1) 
\Text(54.8,11.5)[l]{$G$}
\DashLine(41.7,21.1)(54.3,11.5){3.0} 
\Text(35,0)[b] {diagr.29}
\end{picture} \ 
%  diagram # 42
\begin{picture}(70,80)(0,0)
\Text(15.9,59.5)[r]{$u$}
\ArrowLine(16.3,59.5)(29.0,59.5) 
\Text(35.1,62.1)[b]{$W+$}
\DashArrowLine(29.0,59.5)(41.7,59.5){3.0} 
\Text(54.8,69.1)[l]{$B$}
\ArrowLine(54.3,69.1)(41.7,59.5) 
\Text(40.0,49.9)[r]{$t$}
\ArrowLine(41.7,59.5)(41.7,40.3) 
\Text(54.8,49.9)[l]{$W+$}
\DashArrowLine(41.7,40.3)(54.3,49.9){3.0} 
\Text(54.8,30.7)[l]{$b$}
\ArrowLine(41.7,40.3)(54.3,30.7) 
\Text(27.4,40.3)[r]{$d$}
\ArrowLine(29.0,59.5)(29.0,21.1) 
\Text(15.9,21.1)[r]{$D$}
\ArrowLine(29.0,21.1)(16.3,21.1) 
\DashLine(29.0,21.1)(41.7,21.1){3.0} 
\Text(54.8,11.5)[l]{$G$}
\DashLine(41.7,21.1)(54.3,11.5){3.0} 
\Text(35,0)[b] {diagr.30}
\end{picture} \ 
%  diagram # 43
\begin{picture}(70,80)(0,0)
\Text(15.9,59.5)[r]{$u$}
\ArrowLine(16.3,59.5)(29.0,59.5) 
\Text(35.1,62.1)[b]{$W+$}
\DashArrowLine(29.0,59.5)(41.7,59.5){3.0} 
\Text(54.8,69.1)[l]{$W+$}
\DashArrowLine(41.7,59.5)(54.3,69.1){3.0} 
\Text(41.3,49.9)[r]{$A$}
\DashLine(41.7,59.5)(41.7,40.3){3.0} 
\Text(54.8,49.9)[l]{$b$}
\ArrowLine(41.7,40.3)(54.3,49.9) 
\Text(54.8,30.7)[l]{$B$}
\ArrowLine(54.3,30.7)(41.7,40.3) 
\Text(27.4,40.3)[r]{$d$}
\ArrowLine(29.0,59.5)(29.0,21.1) 
\Text(15.9,21.1)[r]{$D$}
\ArrowLine(29.0,21.1)(16.3,21.1) 
\DashLine(29.0,21.1)(41.7,21.1){3.0} 
\Text(54.8,11.5)[l]{$G$}
\DashLine(41.7,21.1)(54.3,11.5){3.0} 
\Text(35,0)[b] {diagr.31}
\end{picture} \ 
%  diagram # 44
\begin{picture}(70,80)(0,0)
\Text(15.9,59.5)[r]{$u$}
\ArrowLine(16.3,59.5)(29.0,59.5) 
\Text(35.1,62.1)[b]{$W+$}
\DashArrowLine(29.0,59.5)(41.7,59.5){3.0} 
\Text(54.8,69.1)[l]{$W+$}
\DashArrowLine(41.7,59.5)(54.3,69.1){3.0} 
\Text(41.3,49.9)[r]{$H$}
\DashLine(41.7,59.5)(41.7,40.3){1.0}
\Text(54.8,49.9)[l]{$b$}
\ArrowLine(41.7,40.3)(54.3,49.9) 
\Text(54.8,30.7)[l]{$B$}
\ArrowLine(54.3,30.7)(41.7,40.3) 
\Text(27.4,40.3)[r]{$d$}
\ArrowLine(29.0,59.5)(29.0,21.1) 
\Text(15.9,21.1)[r]{$D$}
\ArrowLine(29.0,21.1)(16.3,21.1) 
\DashLine(29.0,21.1)(41.7,21.1){3.0} 
\Text(54.8,11.5)[l]{$G$}
\DashLine(41.7,21.1)(54.3,11.5){3.0} 
\Text(35,0)[b] {diagr.32}
\end{picture} \ 
%  diagram # 45
\begin{picture}(70,80)(0,0)
\Text(15.9,59.5)[r]{$u$}
\ArrowLine(16.3,59.5)(29.0,59.5) 
\Text(35.1,62.1)[b]{$W+$}
\DashArrowLine(29.0,59.5)(41.7,59.5){3.0} 
\Text(54.8,69.1)[l]{$W+$}
\DashArrowLine(41.7,59.5)(54.3,69.1){3.0} 
\Text(41.3,49.9)[r]{$Z$}
\DashLine(41.7,59.5)(41.7,40.3){3.0} 
\Text(54.8,49.9)[l]{$b$}
\ArrowLine(41.7,40.3)(54.3,49.9) 
\Text(54.8,30.7)[l]{$B$}
\ArrowLine(54.3,30.7)(41.7,40.3) 
\Text(27.4,40.3)[r]{$d$}
\ArrowLine(29.0,59.5)(29.0,21.1) 
\Text(15.9,21.1)[r]{$D$}
\ArrowLine(29.0,21.1)(16.3,21.1) 
\DashLine(29.0,21.1)(41.7,21.1){3.0} 
\Text(54.8,11.5)[l]{$G$}
\DashLine(41.7,21.1)(54.3,11.5){3.0} 
\Text(35,0)[b] {diagr.33}
\end{picture} \ 
%  diagram # 46
\begin{picture}(70,80)(0,0)
\Text(15.9,59.5)[r]{$u$}
\ArrowLine(16.3,59.5)(29.0,59.5) 
\DashLine(29.0,59.5)(41.7,59.5){3.0} 
\Text(54.8,69.1)[l]{$W+$}
\DashArrowLine(41.7,59.5)(54.3,69.1){3.0} 
\Text(27.4,49.9)[r]{$d$}
\ArrowLine(29.0,59.5)(29.0,40.3) 
\Text(35.1,41.0)[b]{$A$}
\DashLine(29.0,40.3)(41.7,40.3){3.0} 
\Text(54.8,49.9)[l]{$b$}
\ArrowLine(41.7,40.3)(54.3,49.9) 
\Text(54.8,30.7)[l]{$B$}
\ArrowLine(54.3,30.7)(41.7,40.3) 
\Text(27.4,30.7)[r]{$d$}
\ArrowLine(29.0,40.3)(29.0,21.1) 
\Text(15.9,21.1)[r]{$D$}
\ArrowLine(29.0,21.1)(16.3,21.1) 
\DashLine(29.0,21.1)(41.7,21.1){3.0} 
\Text(54.8,11.5)[l]{$G$}
\DashLine(41.7,21.1)(54.3,11.5){3.0} 
\Text(35,0)[b] {diagr.34}
\end{picture} \ 
%  diagram # 47
\begin{picture}(70,80)(0,0)
\Text(15.9,59.5)[r]{$u$}
\ArrowLine(16.3,59.5)(29.0,59.5) 
\DashLine(29.0,59.5)(41.7,59.5){3.0} 
\Text(54.8,69.1)[l]{$W+$}
\DashArrowLine(41.7,59.5)(54.3,69.1){3.0} 
\Text(27.4,49.9)[r]{$d$}
\ArrowLine(29.0,59.5)(29.0,40.3) 
\Text(15.9,40.3)[r]{$D$}
\ArrowLine(29.0,40.3)(16.3,40.3) 
\Text(35.1,41.0)[b]{$A$}
\DashLine(29.0,40.3)(41.7,40.3){3.0} 
\Text(54.8,49.9)[l]{$B$}
\ArrowLine(54.3,49.9)(41.7,40.3) 
\Text(40.0,30.7)[r]{$b$}
\ArrowLine(41.7,40.3)(41.7,21.1) 
\Text(54.8,30.7)[l]{$b$}
\ArrowLine(41.7,21.1)(54.3,30.7) 
\Text(54.8,11.5)[l]{$G$}
\DashLine(41.7,21.1)(54.3,11.5){3.0} 
\Text(35,0)[b] {diagr.35}
\end{picture} \ 
%  diagram # 48
\begin{picture}(70,80)(0,0)
\Text(15.9,59.5)[r]{$u$}
\ArrowLine(16.3,59.5)(29.0,59.5) 
\DashLine(29.0,59.5)(41.7,59.5){3.0} 
\Text(54.8,69.1)[l]{$W+$}
\DashArrowLine(41.7,59.5)(54.3,69.1){3.0} 
\Text(27.4,49.9)[r]{$d$}
\ArrowLine(29.0,59.5)(29.0,40.3) 
\Text(15.9,40.3)[r]{$D$}
\ArrowLine(29.0,40.3)(16.3,40.3) 
\Text(35.1,41.0)[b]{$A$}
\DashLine(29.0,40.3)(41.7,40.3){3.0} 
\Text(54.8,49.9)[l]{$b$}
\ArrowLine(41.7,40.3)(54.3,49.9) 
\Text(40.0,30.7)[r]{$b$}
\ArrowLine(41.7,21.1)(41.7,40.3) 
\Text(54.8,30.7)[l]{$B$}
\ArrowLine(54.3,30.7)(41.7,21.1) 
\Text(54.8,11.5)[l]{$G$}
\DashLine(41.7,21.1)(54.3,11.5){3.0} 
\Text(35,0)[b] {diagr.36}
\end{picture} \ 
%  diagram # 49
\begin{picture}(70,80)(0,0)
\Text(15.9,59.5)[r]{$u$}
\ArrowLine(16.3,59.5)(29.0,59.5) 
\DashLine(29.0,59.5)(41.7,59.5){3.0} 
\Text(54.8,69.1)[l]{$W+$}
\DashArrowLine(41.7,59.5)(54.3,69.1){3.0} 
\Text(27.4,49.9)[r]{$d$}
\ArrowLine(29.0,59.5)(29.0,40.3) 
\DashLine(29.0,40.3)(41.7,40.3){3.0} 
\Text(54.8,49.9)[l]{$G$}
\DashLine(41.7,40.3)(54.3,49.9){3.0} 
\Text(27.4,30.7)[r]{$d$}
\ArrowLine(29.0,40.3)(29.0,21.1) 
\Text(15.9,21.1)[r]{$D$}
\ArrowLine(29.0,21.1)(16.3,21.1) 
\Text(35.1,21.8)[b]{$A$}
\DashLine(29.0,21.1)(41.7,21.1){3.0} 
\Text(54.8,30.7)[l]{$b$}
\ArrowLine(41.7,21.1)(54.3,30.7) 
\Text(54.8,11.5)[l]{$B$}
\ArrowLine(54.3,11.5)(41.7,21.1) 
\Text(35,0)[b] {diagr.37}
\end{picture} \ 
%  diagram # 50
\begin{picture}(70,80)(0,0)
\Text(15.9,59.5)[r]{$u$}
\ArrowLine(16.3,59.5)(29.0,59.5) 
\DashLine(29.0,59.5)(41.7,59.5){3.0} 
\Text(54.8,69.1)[l]{$W+$}
\DashArrowLine(41.7,59.5)(54.3,69.1){3.0} 
\Text(27.4,49.9)[r]{$d$}
\ArrowLine(29.0,59.5)(29.0,40.3) 
\DashLine(29.0,40.3)(41.7,40.3){3.0} 
\Text(54.8,49.9)[l]{$G$}
\DashLine(41.7,40.3)(54.3,49.9){3.0} 
\Text(27.4,30.7)[r]{$d$}
\ArrowLine(29.0,40.3)(29.0,21.1) 
\Text(15.9,21.1)[r]{$D$}
\ArrowLine(29.0,21.1)(16.3,21.1) 
\Text(35.1,21.8)[b]{$G$}
\DashLine(29.0,21.1)(41.7,21.1){3.0} 
\Text(54.8,30.7)[l]{$b$}
\ArrowLine(41.7,21.1)(54.3,30.7) 
\Text(54.8,11.5)[l]{$B$}
\ArrowLine(54.3,11.5)(41.7,21.1) 
\Text(35,0)[b] {diagr.38}
\end{picture} \ 
%  diagram # 51
\begin{picture}(70,80)(0,0)
\Text(15.9,59.5)[r]{$u$}
\ArrowLine(16.3,59.5)(29.0,59.5) 
\DashLine(29.0,59.5)(41.7,59.5){3.0} 
\Text(54.8,69.1)[l]{$W+$}
\DashArrowLine(41.7,59.5)(54.3,69.1){3.0} 
\Text(27.4,49.9)[r]{$d$}
\ArrowLine(29.0,59.5)(29.0,40.3) 
\DashLine(29.0,40.3)(41.7,40.3){3.0} 
\Text(54.8,49.9)[l]{$G$}
\DashLine(41.7,40.3)(54.3,49.9){3.0} 
\Text(27.4,30.7)[r]{$d$}
\ArrowLine(29.0,40.3)(29.0,21.1) 
\Text(15.9,21.1)[r]{$D$}
\ArrowLine(29.0,21.1)(16.3,21.1) 
\Text(35.1,21.8)[b]{$Z$}
\DashLine(29.0,21.1)(41.7,21.1){3.0} 
\Text(54.8,30.7)[l]{$b$}
\ArrowLine(41.7,21.1)(54.3,30.7) 
\Text(54.8,11.5)[l]{$B$}
\ArrowLine(54.3,11.5)(41.7,21.1) 
\Text(35,0)[b] {diagr.39}
\end{picture} \ 
%  diagram # 52
\begin{picture}(70,80)(0,0)
\Text(15.9,59.5)[r]{$u$}
\ArrowLine(16.3,59.5)(29.0,59.5) 
\DashLine(29.0,59.5)(41.7,59.5){3.0} 
\Text(54.8,69.1)[l]{$W+$}
\DashArrowLine(41.7,59.5)(54.3,69.1){3.0} 
\Text(27.4,49.9)[r]{$d$}
\ArrowLine(29.0,59.5)(29.0,40.3) 
\Text(35.1,41.0)[b]{$G$}
\DashLine(29.0,40.3)(41.7,40.3){3.0} 
\Text(54.8,49.9)[l]{$b$}
\ArrowLine(41.7,40.3)(54.3,49.9) 
\Text(54.8,30.7)[l]{$B$}
\ArrowLine(54.3,30.7)(41.7,40.3) 
\Text(27.4,30.7)[r]{$d$}
\ArrowLine(29.0,40.3)(29.0,21.1) 
\Text(15.9,21.1)[r]{$D$}
\ArrowLine(29.0,21.1)(16.3,21.1) 
\DashLine(29.0,21.1)(41.7,21.1){3.0} 
\Text(54.8,11.5)[l]{$G$}
\DashLine(41.7,21.1)(54.3,11.5){3.0} 
\Text(35,0)[b] {diagr.40}
\end{picture} \ 
%  diagram # 53
\begin{picture}(70,80)(0,0)
\Text(15.9,59.5)[r]{$u$}
\ArrowLine(16.3,59.5)(29.0,59.5) 
\DashLine(29.0,59.5)(41.7,59.5){3.0} 
\Text(54.8,69.1)[l]{$W+$}
\DashArrowLine(41.7,59.5)(54.3,69.1){3.0} 
\Text(27.4,49.9)[r]{$d$}
\ArrowLine(29.0,59.5)(29.0,40.3) 
\Text(15.9,40.3)[r]{$D$}
\ArrowLine(29.0,40.3)(16.3,40.3) 
\Text(35.1,41.0)[b]{$G$}
\DashLine(29.0,40.3)(41.7,40.3){3.0} 
\Text(54.8,49.9)[l]{$G$}
\DashLine(41.7,40.3)(54.3,49.9){3.0} 
\Text(41.3,30.7)[r]{$G$}
\DashLine(41.7,40.3)(41.7,21.1){3.0} 
\Text(54.8,30.7)[l]{$b$}
\ArrowLine(41.7,21.1)(54.3,30.7) 
\Text(54.8,11.5)[l]{$B$}
\ArrowLine(54.3,11.5)(41.7,21.1) 
\Text(35,0)[b] {diagr.41}
\end{picture} \ 
%  diagram # 54
\begin{picture}(70,80)(0,0)
\Text(15.9,59.5)[r]{$u$}
\ArrowLine(16.3,59.5)(29.0,59.5) 
\DashLine(29.0,59.5)(41.7,59.5){3.0} 
\Text(54.8,69.1)[l]{$W+$}
\DashArrowLine(41.7,59.5)(54.3,69.1){3.0} 
\Text(27.4,49.9)[r]{$d$}
\ArrowLine(29.0,59.5)(29.0,40.3) 
\Text(15.9,40.3)[r]{$D$}
\ArrowLine(29.0,40.3)(16.3,40.3) 
\Text(35.1,41.0)[b]{$G$}
\DashLine(29.0,40.3)(41.7,40.3){3.0} 
\Text(54.8,49.9)[l]{$B$}
\ArrowLine(54.3,49.9)(41.7,40.3) 
\Text(40.0,30.7)[r]{$b$}
\ArrowLine(41.7,40.3)(41.7,21.1) 
\Text(54.8,30.7)[l]{$b$}
\ArrowLine(41.7,21.1)(54.3,30.7) 
\Text(54.8,11.5)[l]{$G$}
\DashLine(41.7,21.1)(54.3,11.5){3.0} 
\Text(35,0)[b] {diagr.42}
\end{picture} \ 
%  diagram # 55
\begin{picture}(70,80)(0,0)
\Text(15.9,59.5)[r]{$u$}
\ArrowLine(16.3,59.5)(29.0,59.5) 
\DashLine(29.0,59.5)(41.7,59.5){3.0} 
\Text(54.8,69.1)[l]{$W+$}
\DashArrowLine(41.7,59.5)(54.3,69.1){3.0} 
\Text(27.4,49.9)[r]{$d$}
\ArrowLine(29.0,59.5)(29.0,40.3) 
\Text(15.9,40.3)[r]{$D$}
\ArrowLine(29.0,40.3)(16.3,40.3) 
\Text(35.1,41.0)[b]{$G$}
\DashLine(29.0,40.3)(41.7,40.3){3.0} 
\Text(54.8,49.9)[l]{$b$}
\ArrowLine(41.7,40.3)(54.3,49.9) 
\Text(40.0,30.7)[r]{$b$}
\ArrowLine(41.7,21.1)(41.7,40.3) 
\Text(54.8,30.7)[l]{$B$}
\ArrowLine(54.3,30.7)(41.7,21.1) 
\Text(54.8,11.5)[l]{$G$}
\DashLine(41.7,21.1)(54.3,11.5){3.0} 
\Text(35,0)[b] {diagr.43}
\end{picture} \ 
%  diagram # 56
\begin{picture}(70,80)(0,0)
\Text(15.9,59.5)[r]{$u$}
\ArrowLine(16.3,59.5)(29.0,59.5) 
\DashLine(29.0,59.5)(41.7,59.5){3.0} 
\Text(54.8,69.1)[l]{$W+$}
\DashArrowLine(41.7,59.5)(54.3,69.1){3.0} 
\Text(27.4,49.9)[r]{$d$}
\ArrowLine(29.0,59.5)(29.0,40.3) 
\Text(15.9,40.3)[r]{$D$}
\ArrowLine(29.0,40.3)(16.3,40.3) 
\Text(35.1,41.0)[b]{$Z$}
\DashLine(29.0,40.3)(41.7,40.3){3.0} 
\Text(54.8,49.9)[l]{$B$}
\ArrowLine(54.3,49.9)(41.7,40.3) 
\Text(40.0,30.7)[r]{$b$}
\ArrowLine(41.7,40.3)(41.7,21.1) 
\Text(54.8,30.7)[l]{$b$}
\ArrowLine(41.7,21.1)(54.3,30.7) 
\Text(54.8,11.5)[l]{$G$}
\DashLine(41.7,21.1)(54.3,11.5){3.0} 
\Text(35,0)[b] {diagr.44}
\end{picture} \ 
%  diagram # 57
\begin{picture}(70,80)(0,0)
\Text(15.9,59.5)[r]{$u$}
\ArrowLine(16.3,59.5)(29.0,59.5) 
\DashLine(29.0,59.5)(41.7,59.5){3.0} 
\Text(54.8,69.1)[l]{$W+$}
\DashArrowLine(41.7,59.5)(54.3,69.1){3.0} 
\Text(27.4,49.9)[r]{$d$}
\ArrowLine(29.0,59.5)(29.0,40.3) 
\Text(15.9,40.3)[r]{$D$}
\ArrowLine(29.0,40.3)(16.3,40.3) 
\Text(35.1,41.0)[b]{$Z$}
\DashLine(29.0,40.3)(41.7,40.3){3.0} 
\Text(54.8,49.9)[l]{$b$}
\ArrowLine(41.7,40.3)(54.3,49.9) 
\Text(40.0,30.7)[r]{$b$}
\ArrowLine(41.7,21.1)(41.7,40.3) 
\Text(54.8,30.7)[l]{$B$}
\ArrowLine(54.3,30.7)(41.7,21.1) 
\Text(54.8,11.5)[l]{$G$}
\DashLine(41.7,21.1)(54.3,11.5){3.0} 
\Text(35,0)[b] {diagr.45}
\end{picture} \ 
%  diagram # 58
\begin{picture}(70,80)(0,0)
\Text(15.9,59.5)[r]{$u$}
\ArrowLine(16.3,59.5)(29.0,59.5) 
\DashLine(29.0,59.5)(41.7,59.5){3.0} 
\Text(54.8,69.1)[l]{$W+$}
\DashArrowLine(41.7,59.5)(54.3,69.1){3.0} 
\Text(27.4,49.9)[r]{$d$}
\ArrowLine(29.0,59.5)(29.0,40.3) 
\Text(35.1,41.0)[b]{$Z$}
\DashLine(29.0,40.3)(41.7,40.3){3.0} 
\Text(54.8,49.9)[l]{$b$}
\ArrowLine(41.7,40.3)(54.3,49.9) 
\Text(54.8,30.7)[l]{$B$}
\ArrowLine(54.3,30.7)(41.7,40.3) 
\Text(27.4,30.7)[r]{$d$}
\ArrowLine(29.0,40.3)(29.0,21.1) 
\Text(15.9,21.1)[r]{$D$}
\ArrowLine(29.0,21.1)(16.3,21.1) 
\DashLine(29.0,21.1)(41.7,21.1){3.0} 
\Text(54.8,11.5)[l]{$G$}
\DashLine(41.7,21.1)(54.3,11.5){3.0} 
\Text(35,0)[b] {diagr.46}
\end{picture} \ 
%  diagram # 71
\begin{picture}(70,80)(0,0)
\Text(15.9,59.5)[r]{$u$}
\ArrowLine(16.3,59.5)(29.0,59.5) 
\Text(35.1,60.2)[b]{$Z$}
\DashLine(29.0,59.5)(41.7,59.5){3.0} 
\Text(54.8,69.1)[l]{$B$}
\ArrowLine(54.3,69.1)(41.7,59.5) 
\Text(40.0,49.9)[r]{$b$}
\ArrowLine(41.7,59.5)(41.7,40.3) 
\Text(54.8,49.9)[l]{$b$}
\ArrowLine(41.7,40.3)(54.3,49.9) 
\Text(54.8,30.7)[l]{$G$}
\DashLine(41.7,40.3)(54.3,30.7){3.0} 
\Text(27.4,40.3)[r]{$u$}
\ArrowLine(29.0,59.5)(29.0,21.1) 
\Text(15.9,21.1)[r]{$D$}
\ArrowLine(29.0,21.1)(16.3,21.1) 
\DashLine(29.0,21.1)(41.7,21.1){3.0} 
\Text(54.8,11.5)[l]{$W+$}
\DashArrowLine(41.7,21.1)(54.3,11.5){3.0} 
\Text(35,0)[b] {diagr.47}
\end{picture} \ 
%  diagram # 72
\begin{picture}(70,80)(0,0)
\Text(15.9,59.5)[r]{$u$}
\ArrowLine(16.3,59.5)(29.0,59.5) 
\Text(35.1,60.2)[b]{$Z$}
\DashLine(29.0,59.5)(41.7,59.5){3.0} 
\Text(54.8,69.1)[l]{$b$}
\ArrowLine(41.7,59.5)(54.3,69.1) 
\Text(40.0,49.9)[r]{$b$}
\ArrowLine(41.7,40.3)(41.7,59.5) 
\Text(54.8,49.9)[l]{$B$}
\ArrowLine(54.3,49.9)(41.7,40.3) 
\Text(54.8,30.7)[l]{$G$}
\DashLine(41.7,40.3)(54.3,30.7){3.0} 
\Text(27.4,40.3)[r]{$u$}
\ArrowLine(29.0,59.5)(29.0,21.1) 
\Text(15.9,21.1)[r]{$D$}
\ArrowLine(29.0,21.1)(16.3,21.1) 
\DashLine(29.0,21.1)(41.7,21.1){3.0} 
\Text(54.8,11.5)[l]{$W+$}
\DashArrowLine(41.7,21.1)(54.3,11.5){3.0} 
\Text(35,0)[b] {diagr.48}
\end{picture} \ 
%  diagram # 73
\begin{picture}(70,80)(0,0)
\Text(15.9,59.5)[r]{$u$}
\ArrowLine(16.3,59.5)(29.0,59.5) 
\Text(35.1,60.2)[b]{$Z$}
\DashLine(29.0,59.5)(41.7,59.5){3.0} 
\Text(54.8,69.1)[l]{$b$}
\ArrowLine(41.7,59.5)(54.3,69.1) 
\Text(54.8,49.9)[l]{$B$}
\ArrowLine(54.3,49.9)(41.7,59.5) 
\Text(27.4,49.9)[r]{$u$}
\ArrowLine(29.0,59.5)(29.0,40.3) 
\DashLine(29.0,40.3)(41.7,40.3){3.0} 
\Text(54.8,30.7)[l]{$G$}
\DashLine(41.7,40.3)(54.3,30.7){3.0} 
\Text(27.4,30.7)[r]{$u$}
\ArrowLine(29.0,40.3)(29.0,21.1) 
\Text(15.9,21.1)[r]{$D$}
\ArrowLine(29.0,21.1)(16.3,21.1) 
\DashLine(29.0,21.1)(41.7,21.1){3.0} 
\Text(54.8,11.5)[l]{$W+$}
\DashArrowLine(41.7,21.1)(54.3,11.5){3.0} 
\Text(35,0)[b] {diagr.49}
\end{picture} \ 
%  diagram # 74
\begin{picture}(70,80)(0,0)
\Text(15.9,59.5)[r]{$u$}
\ArrowLine(16.3,59.5)(29.0,59.5) 
\Text(35.1,60.2)[b]{$Z$}
\DashLine(29.0,59.5)(41.7,59.5){3.0} 
\Text(54.8,69.1)[l]{$b$}
\ArrowLine(41.7,59.5)(54.3,69.1) 
\Text(54.8,49.9)[l]{$B$}
\ArrowLine(54.3,49.9)(41.7,59.5) 
\Text(27.4,49.9)[r]{$u$}
\ArrowLine(29.0,59.5)(29.0,40.3) 
\DashLine(29.0,40.3)(41.7,40.3){3.0} 
\Text(54.8,30.7)[l]{$W+$}
\DashArrowLine(41.7,40.3)(54.3,30.7){3.0} 
\Text(27.4,30.7)[r]{$d$}
\ArrowLine(29.0,40.3)(29.0,21.1) 
\Text(15.9,21.1)[r]{$D$}
\ArrowLine(29.0,21.1)(16.3,21.1) 
\DashLine(29.0,21.1)(41.7,21.1){3.0} 
\Text(54.8,11.5)[l]{$G$}
\DashLine(41.7,21.1)(54.3,11.5){3.0} 
\Text(35,0)[b] {diagr.50}
\end{picture}
}
  \end{center}
  \caption{\label{fig:feyn-jbbw}%
    Feynman diagrams for the process $u\bar{d}\to gb\bar{b}W$.}
\end{figure}
The Feynman diagrams for process~(\ref{eq:pp->bbW}) at parton level are
shown in figure~\ref{fig:feyn-bbw} and the Feynman diagrams for
process~$u\bar{d}\to b\bar{b}W+\text{jet}$ as a representative of the
processes contributing to the process~(\ref{eq:pp->bbWj}) at parton level
are shown in figure~\ref{fig:feyn-jbbw}.  Parton processes with gluons
in the initial state are not shown, but are included in the
calculation.  The diagrams include the top signal, Higgs contribution,
QCD diagrams with gluons in the intermediate state and several other
electroweak diagrams which have to be taken into account for an
accurate calculation of the rate of very hard processes.  Here, the
Higgs contribution is part of the background to the single top
production.  In our calculation, we have assumed a light Higgs with a
mass in the range 80--120~GeV as a worst case scenario for the single
top signal, since the Higgs rate drops rapidly with rising Higgs
mass~\cite{higgs}.  Even in this case, the Higgs
contributions will turn out to very small in the phase space regions
which will be important for single top production.  

All the calculations have been performed with the computer program
Comp\-HEP~\cite{comphep}, including the proper mapping of singularities
and smoothing in singular variables~\cite{pukhov}.  We have used the
NLO CTEQ4M parametrization of parton distribution
functions~\cite{cteq}.  For the quark induced
processes~(\ref{eq:pp->bbW}), we have chosen the QCD factorization scale
to be~$M_t$.  This choice is dictated by the fact that we are
selecting a kinematical region where the two quarks annihilate into a
state close to the top quark mass shell.  For processes involving
$W$-gluon fusion~(\ref{eq:pp->bbWj}), the choice of scale is more subtle,
as has been pointed out by~\cite{willenbrock}.  Therefore, we have
pragmatically fixed the scale by matching our LO cross sections to the
NLO results of~\cite{willenbrock}.  This procedure leads us to a
factorization scale of~$Q^2\approx(M_t/2)^2$.  The fact that this
scale is reasonably close to the top quark mass shows that the
corrections are not very large and serves as a \emph{a posteriori}
justification of the pragmatical procedure.  Therefore we have taken
into account the important parts of the NLO corrections in the hard
kinematical region which we are interested in.  Finally, we notice
that the requirement of a double $b$-tag in the hard kinematical
region under consideration suppresses the contributions from the
processes with the $b$-quark in the initial state.  Therefore this
source of theoretical uncertainties~\cite{willenbrock} for the signal
is absent in this case.

%%%%%%%%%%%%%%%%%%%%%%%%%%%%%%%%%%%%%%%%%%%%%%%%%%%%%%%%%%%%%%%%%%%%%%%%
\section{Anomalous $Wtb$ Couplings}
In the model independent effective Lagrangian
approach~\cite{buchmueller-hagiwara-gounaris1} seven anomalous CP
conserving operators of dimension six contribute to the $Wtb$~vertex
with four independent form factors (cf.~\cite{whisnant} for explicit
expressions).  In this paper, we do not attempt a simultaneous
analysis of all seven operators.  Instead we study two anomalous
operators of the magnetic type as an example in order to explore the
potential of the colliders.
In fact, the $V-A$ coupling is as in the SM with the coupling $V_{tb}$
very close to unity, as required by present data~\cite{pdg}.
A possible $V+A$ form factor is severely constrained~\cite{whisnant} by
the CLEO $b\to s\gamma$ data~\cite{cleo} on a level which is stronger
than expected even at high energy $\gamma e$ colliders~\cite{boos1}.
This leaves us with the remaining two magnetic form factors and we have
studied them in the processes~(\ref{eq:pp->bbW(j)}).

As before~\cite{boos1}, we have adopted the notation for the
Lagrangian in unitary gauge from~\cite{kane}:
\begin{equation}
\label{eq:lagrangian_anom}
   \mathcal{L} = \frac{g}{\sqrt{2}}
    \left[ W^-_\nu \bar{b}\gamma_\mu P_- t
     - \frac{1}{2M_W} W^-_{\mu\nu}
           \bar{b}\sigma^{\mu\nu} (F_2^L P_- + F_2^R P_+) t \right]
     + \text{h.\,c.} 
\end{equation}
with $W^\pm_{\mu\nu} = D_\mu W^\pm_\nu - D_\nu W^\pm_{\mu}$,
$D_\mu=\partial_\mu-ieA_\mu$,
$\sigma^{\mu\nu}=i/2[\gamma_\mu,\gamma_\nu]$
and~$P_\pm=(1\pm\gamma_5)/2$.  The
couplings~$F_2^L$ and~$F_2^R$ are proportional to the coefficients
$C_{tW\Phi}$ and $C_{bW\Phi}$ of the effective Lagrangian
\begin{equation}
\label{eq:F2LR}
  F_2^{L(R)} = \frac{C_{t(b)W \Phi}}{\Lambda^2}
               \frac{\sqrt{2} v M_W}{g} 
\end{equation}
The resulting Feynman rules (cf.~the appendix of~\cite{boos1}) have
been implemented in Comp\-HEP.

%%%%%%%%%%%%%%%%%%%%%%%%%%%%%%%%%%%%%%%%%%%%%%%%%%%%%%%%%%%%%%%%%%%%%%%%
\section{%
  Sensitive Variables, Background Suppression and the
  Structure of Singularities in Feynman Diagrams}

As always, the correct mapping of the singularities of the Feynman
diagrams is absolutely crucial in order to achieve numerically stable
results in the MC integration over phase space.  In this section,
however, we will focus on a different aspect of the singularities,
that has impact on physics analysises.

Unless particular cuts are applied, most of the contributions to the
rate of any given process come from phase space regions close to the
singularities.  Indeed, this simple observation forms the basis for
most of our intuition for finding selection cuts to enhance the
signal.  In many cases, it is however also possible to reverse this
argument and use the singularities to systematically devise cuts for
background suppression.  This approach requires an analysis of
\emph{all} Feynman diagrams contributing to the process, signal and
background.

Shifting the focus from signal enhancement to background suppression
in this manner is useful for processes with a high rate of both signal
and background that can afford to lose some rate.  We will demonstrate
that single top production falls into this class.

The general procedure compares the set~$\mathcal{S}$ of variables with
singularities from all signal diagrams with the same set~$\mathcal{B}$
{}from all background diagrams, reducible and irreducible.
If~$\mathcal{B}\setminus\mathcal{S}\not=\emptyset$, i.\,e.~there are
variables with singularities in the background diagrams which are
regular in all signal diagrams, then it is obvious that singularities
in these variables should be cut out as strongly as possible.  It
turns out that the number of different singular variables is very
limited in the cases of practical interest and a general classification
allows recommendations for choosing sensitive variables.  The
application of this approach to neural net methods will be discussed
in~\cite{boos-dudko}.

Shifting our attention back to the special case of single top
production, we note that the single signal diagram
for~(\ref{eq:pp->bbW}) has only one singular variable: the invariant
mass~$M_{Wb}$ of the top decay products near the top
pole~$M_{Wb}=M_t$.  These contributions have to be kept, of course.

In the background diagrams, the only $s$-channel singularities are in
the invariant mass~$M_{b\bar b}$ of the $b\bar b$ pair at $0$, $M_Z$,
and~$M_H$ from the coupling to neutral vector bosons and Higgs
particles.  Since the CKM matrix element~$V_{ub}$ is tiny, we can
ignore the multiperipheral diagrams with the $Wub$-coupling and we
only have to consider the $t$-channel variables~$t_{u\to b\bar b} =
t_{\bar d\to W}$ and~$t_{\bar d\to b\bar b} = t_{u\to W}$.
Unfortunately, the $t$-channel variables are not directly observable in hadron
collisions.  We can, however, use the corresponding transverse momentum
as a surrogate.  In this case, these are the $P^t$ of the $b\bar b$-pair
or, equivalently, of the $W$-boson.

{}From this simple argument, we conclude that the invariant
mass~$M_{b\bar b}$ and the transverse momentum $P^t_W$ are the most
effective variables for expressing cuts for the
process~(\ref{eq:pp->bbW}).

Analogous considerations for the diagrams in
figure~\ref{fig:feyn-jbbw} leads to the variables~$M_{b\bar b}$,
$P^t_{b\bar b}$ and~$P^t_W$ (the latter are now no longer equivalent)
for the process~(\ref{eq:pp->bbWj}).  Here, the transverse momentum
distributions of single jets~$P^t_b$ and~$P^t_q$ are problematic,
because the same singularity occurs for both signal and background
diagrams and the signal and the background will have similar shapes.
Therefore cuts on these variables must be defined with discretion to
achieve a balance between the competing goals of good jet
identification and high signal rate.

Of course, there are more variables that can have different
distributions for the signal and background, but which are not
directly related to the singularities of Feynman diagrams.  One such
variable is the partonic center of mass energy~$\hat{s}$.  The
difference is caused here by different thresholds for the signal and
the backgrounds.

%%%%%%%%%%%%%%%%%%%%%%%%%%%%%%%%%%%%%%%%%%%%%%%%%%%%%%%%%%%%%%%%%%%%%%%%
\section{Distributions and numerical results}

\begin{figure}
  \begin{center}
    \includegraphics[width=\textwidth]{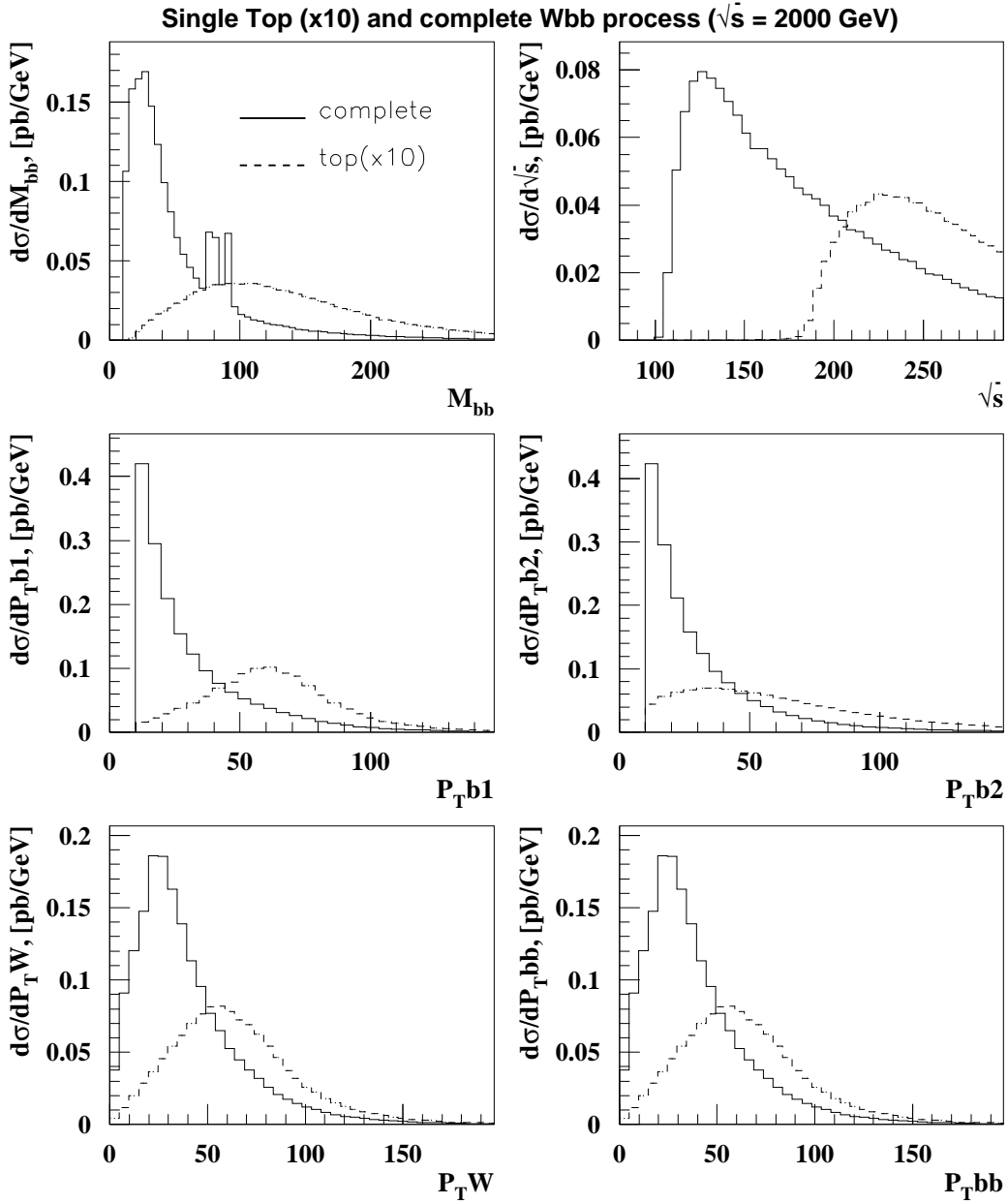}
  \end{center}
  \caption{\label{fig:dist_bbw_TEV}%
    Distributions for invariant masses and transverse momenta for the
    process $p\bar p\to  b\bar{b}W$ at Tevatron using the soft cuts
    in~(\ref{eq:soft-cuts@TeV}).}
\end{figure}
\begin{figure}
  \begin{center}
    \includegraphics[width=\textwidth]{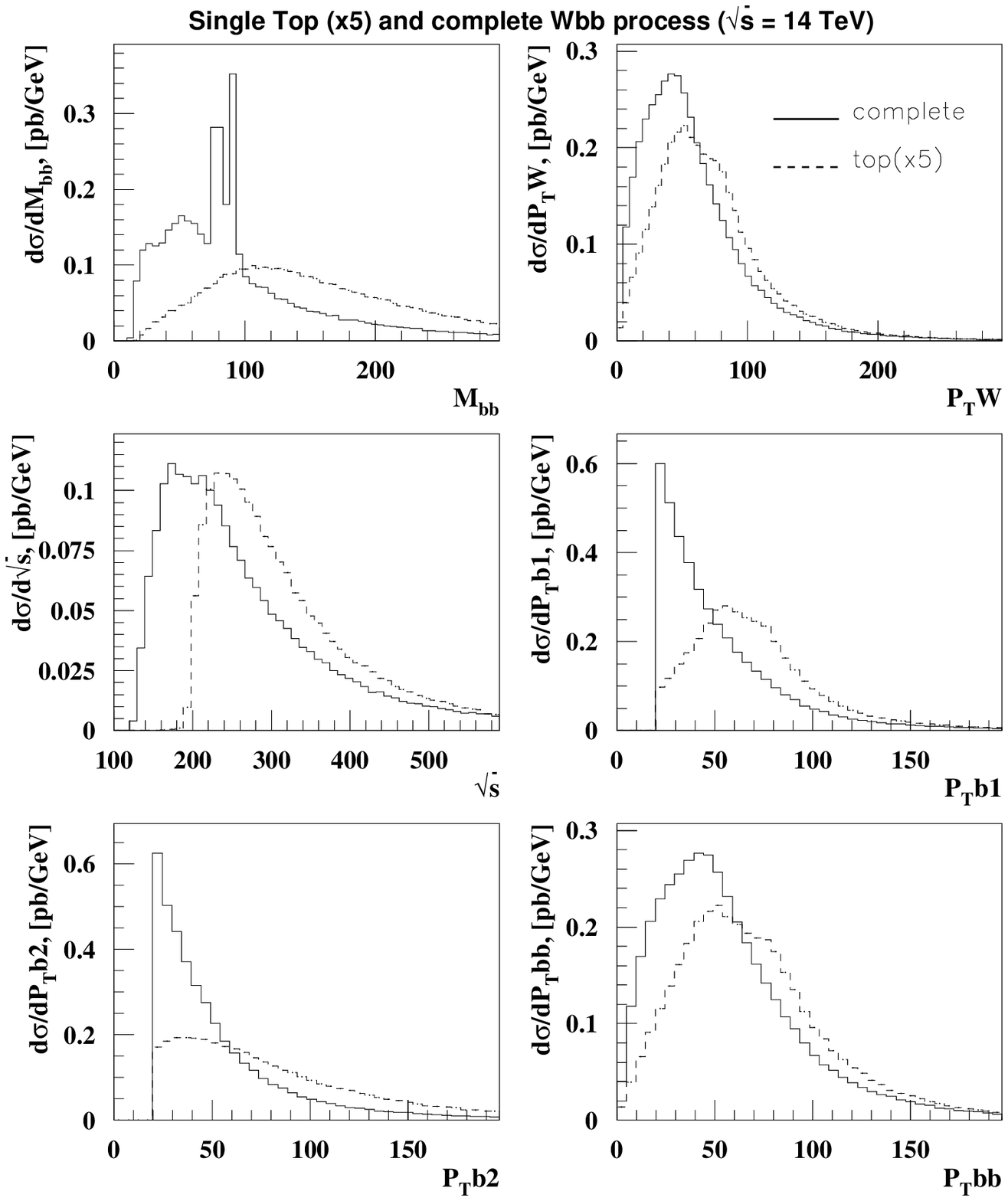}
  \end{center}
  \caption{\label{fig:dist_bbw_LHC}%
    Distributions for invariant masses and transverse momenta for the
    process $p\bar p\to  b\bar{b}W$ at LHC using the soft cuts
    in~(\ref{eq:soft-cuts@LHC}).}
\end{figure}
\begin{figure}
  \begin{center}
    \includegraphics[width=\textwidth]{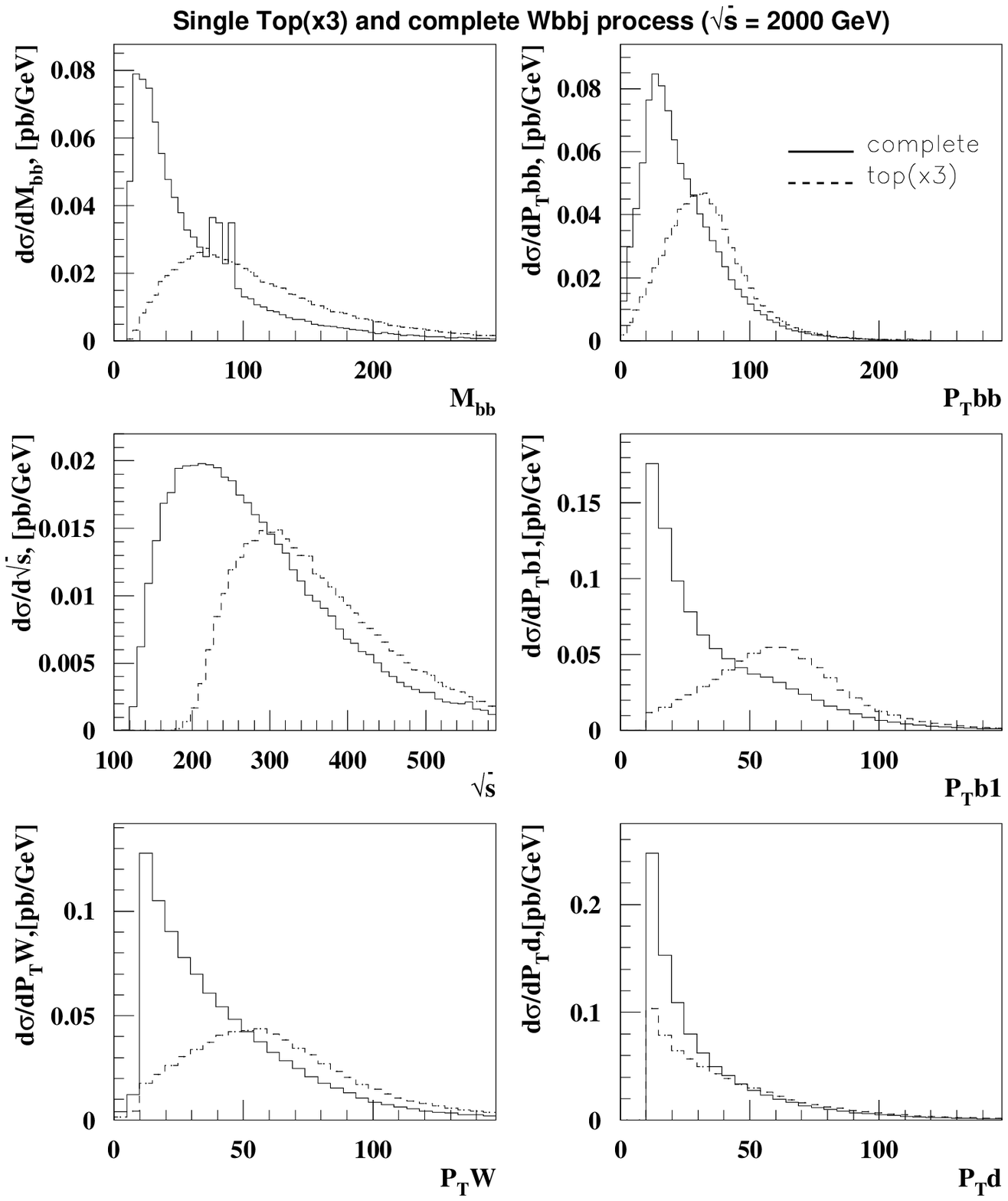}
  \end{center}
  \caption{\label{fig:dist_jbbw_TEV}%
    Distributions for invariant masses and transverse momenta for the
    process $pp\to jb\bar{b}W$ at Tevatron using the soft cuts
    in~(\ref{eq:soft-cuts@TeV}).}
\end{figure}
\begin{figure}
  \begin{center}
    \includegraphics[width=\textwidth]{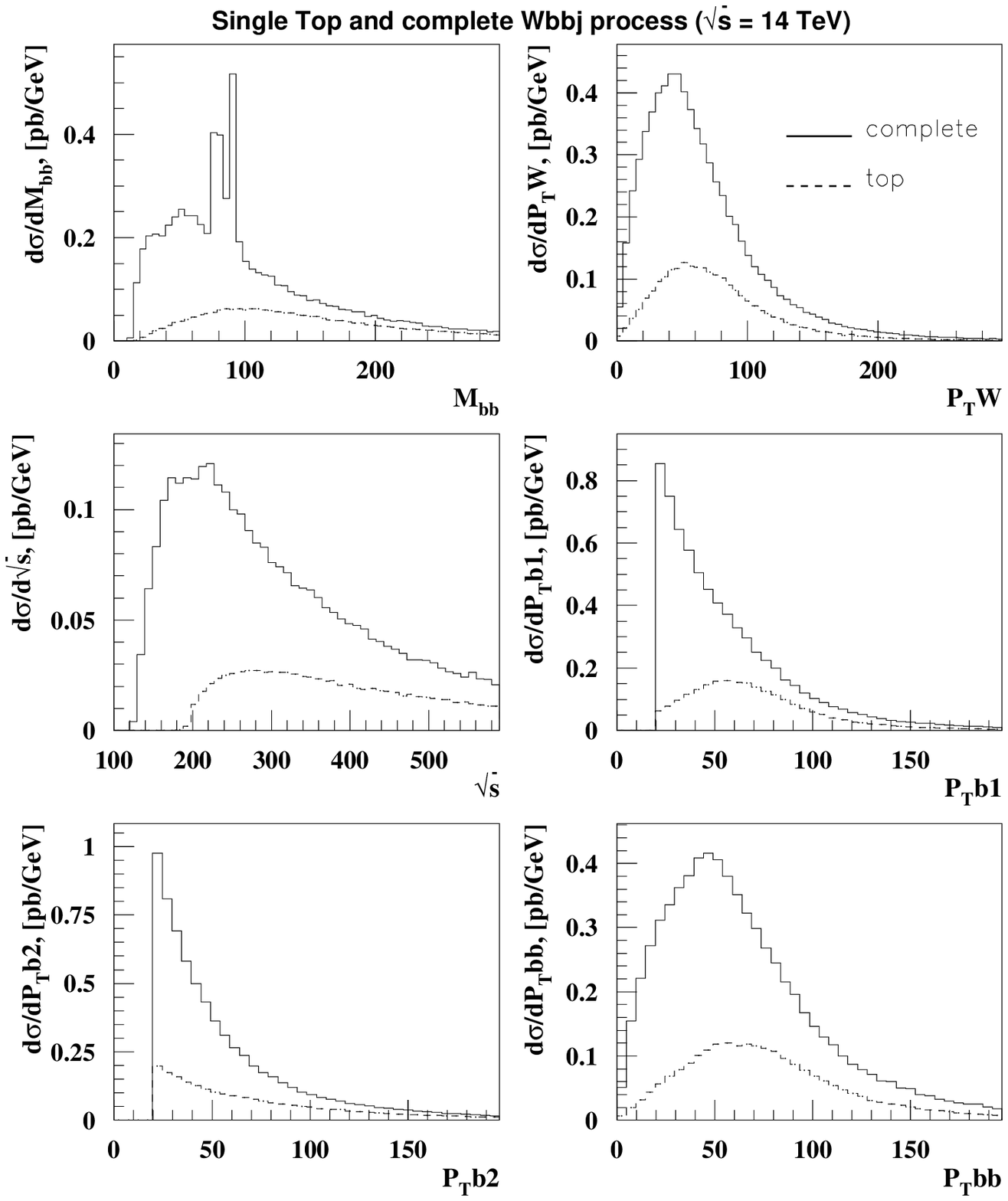}
  \end{center}
  \caption{\label{fig:dist_jbbw_LHC}%
    Distributions for invariant masses and transverse momenta for the
    process $pp\to  jb\bar{b}W$ at LHC using the soft cuts
    in~(\ref{eq:soft-cuts@LHC}).}
\end{figure}
To illustrate the kinematical properties of the
processes~(\ref{eq:pp->bbW(j)}) we show in
figures~\ref{fig:dist_bbw_TEV}, \ref{fig:dist_bbw_LHC},
\ref{fig:dist_jbbw_TEV}, and~\ref{fig:dist_jbbw_LHC} several
distributions on the variables discussed above with soft initial
cuts on the jet $P^t_j$, jet rapidity and jet cone size~$\Delta
R_{jj(ej)}$
\begin{subequations}
\begin{align}
\label{eq:soft-cuts@TeV}
    P^t_j             &> 10 \mathop{\textrm{GeV}}, & 
    |\eta_j|          &<  2.5,                     &
    \Delta R_{jj(ej)} &>  0.5
  &\bigr\}& && \text{Tevatron} \\
\label{eq:soft-cuts@LHC}
    P^t_j             &> 20 \mathop{\textrm{GeV}}, &
    |\eta_j|          &<  3,                       &
    \Delta R_{jj(ej)} &>  0.5
  &\bigr\}& && \text{LHC}
\end{align}
\end{subequations}
The figures allow to compare the distributions from the single top
part only with those from the complete set of the SM diagrams. The
notations~$b_1$ and~$b_2$ refer to the $b$-jets with the larger and
smaller~$P^t$ respectively.  To make the contribution of the single
top more visible in the figures, we have scaled the rate by
appropriate factors as indicated.  As we have expected from the
analysis of the singularities, the distributions are significantly
different for single top and background contributions.

The additional power of momentum in the amomalous
couplings~(\ref{eq:lagrangian_anom}) will cause
a deviation from the~SM prediction that rises with energy and
transversal momentum.  However, since the rate falls off quickly
with~$P^t$, the optimal cuts must not be too strong in order to
conserve rate.

The optimized cuts turn out to be different for the Tevatron and
the~LHC, as well as for processes with two $b$-jets and 
processes with two $b$-jets and one light quark or gluon jet on the
other hand.  For the process~(\ref{eq:pp->bbW}) we find
%%% align seems to interfere with aligned (AMS-LaTeX bug???)
\begin{subequations}
\label{eq:cuts:optimized}
\begin{eqnarray}
\label{eq:cuts:pp->bbW@TeV}
    \left.
      \begin{aligned}
        P^t_{b_1}            &>&  30 \mathop{\textrm{GeV}}, &\quad&
        P^t_{b_2}            &>&  20 \mathop{\textrm{GeV}}, \\
        M_{b\bar b}          &>& 100 \mathop{\textrm{GeV}}, &\quad&
        P^t_{b\bar b}, P^t_W &>&  30 \mathop{\textrm{GeV}}\hphantom{,}
      \end{aligned}
    \right\} && \text{Tevatron} \\
\label{eq:cuts:pp->bbW@LHC}
    \left.
      \begin{aligned}
        P^t_{b_1}            &>&  50 \mathop{\textrm{GeV}}, &\quad&
        P^t_{b_2}            &>&  20 \mathop{\textrm{GeV}}, \\
        M_{b\bar b}          &>& 100 \mathop{\textrm{GeV}}, &\quad&
        P^t_{b\bar b}, P^t_W &>& 100 \mathop{\textrm{GeV}}\hphantom{,}
      \end{aligned}
    \right\} && \text{LHC}
\end{eqnarray}
and for process~(\ref{eq:pp->bbWj})
\begin{eqnarray}
\label{eq:cuts:pp->bbWj@TeV}
    \left.
      \begin{aligned}
        P^t_{b_1}            &>&  40 \mathop{\textrm{GeV}}, &\quad&
        P^t_{b_2}, P^t_j     &>&  20 \mathop{\textrm{GeV}}, &\quad&\\
        M_{b\bar b}          &>&  40 \mathop{\textrm{GeV}}, &\quad&
        P^t_{b\bar b}        &>&  30 \mathop{\textrm{GeV}}, &\quad&
        P^t_W                &>&  20 \mathop{\textrm{GeV}}\hphantom{,}
      \end{aligned}
    \right\} && \text{Tevatron} \\
\label{eq:cuts:pp->bbWj@LHC}
    \left.
      \begin{aligned}
        P^t_{b_1}            &>&  50 \mathop{\textrm{GeV}}, &\quad&
        P^t_{b_2}, P^t_j     &>&  20 \mathop{\textrm{GeV}}, &\quad&\\
        M_{b\bar b}          &>& 100 \mathop{\textrm{GeV}}, &\quad&
        P^t_{b\bar b}        &>& 100 \mathop{\textrm{GeV}}, &\quad&
        P^t_W                &>&  30 \mathop{\textrm{GeV}}\hphantom{,}
      \end{aligned}
    \right\} && \text{LHC}
\end{eqnarray}
\end{subequations}
As we will see below, it is of crucial importance to use both
processes~(\ref{eq:pp->bbW}) and~(\ref{eq:pp->bbWj}) for establishing
limits on anomalous couplings, in particular at the LHC.  In order to
demonstrate the effect of the cuts, we have given the cross sections
for several sub-processes at both colliders in table~\ref{tb:crs1}.
We stress that the rates of single top and single anti-top production
differ together with their corresponding backgrounds at the
$pp$-collider LHC, while they are equal at the $p\bar p$-collider
Tevatron.

\begin{table}
\begin{center}
\begin{tabular}{|c|c|c|c|c|} \hline
      Process  &\multicolumn{2}{|c|}{Tevatron}
               &\multicolumn{2}{|c|}{LHC}                \\
               &\multicolumn{2}{|c|}{$\sigma/\text{pb}$}
               &\multicolumn{2}{|c|}{$\sigma/\text{pb}$} \\\hline\hline
   \multicolumn{1}{|l|}{$u\bar{d}\to W^+b\bar{b}$}
               & soft         & optimized  & soft         & optimized \\
   \multicolumn{1}{|l|}{\quad/ $\bar{u}d\to W^-b\bar{b}$}
               & cuts~(\ref{eq:soft-cuts@TeV})
                              & cuts~(\ref{eq:cuts:pp->bbW@TeV})
                                           & cuts~(\ref{eq:soft-cuts@LHC})
                                           & cuts~(\ref{eq:cuts:pp->bbW@LHC}) \\\hline
   complete    & 8.1          &     0.68   &  16.6  / 10.4 &  3.8 / 2.4 \\       
   single top  &  0.57        &     0.30   &    3.2 /  1.8 &  1.7 / 0.9 \\\hline\hline
   \multicolumn{1}{|l|}{$ug\to dW^+b\bar{b}$}
               & soft         & optimized  & soft         & optimized \\
   \multicolumn{1}{|l|}{\quad/ $\bar{u}g\to \bar{d}W^-b\bar{b}$}
               & cuts~(\ref{eq:soft-cuts@TeV})
                              & cuts~(\ref{eq:cuts:pp->bbWj@TeV})
                                           & cuts~(\ref{eq:soft-cuts@LHC})
                                           & cuts~(\ref{eq:cuts:pp->bbWj@LHC}) \\\hline
   complete    & 1.4          &    0.32    &   28.4 / 5.8 &  9.6 / 1.8 \\      
   single top  & 0.42         &   0.27     &   18.0 / 2.0 &  7.8 / 1.5 \\\hline\hline
   \multicolumn{1}{|l|}{$u\bar{d}\to gW^+b\bar{b}$}
               & soft         & optimized  & soft         & optimized \\
   \multicolumn{1}{|l|}{\quad/ $\bar{u}d\to gW^-b\bar{b}$}
               & cuts~(\ref{eq:soft-cuts@TeV})
                              & cuts~(\ref{eq:cuts:pp->bbWj@TeV})
                                           & cuts~(\ref{eq:soft-cuts@LHC})
                                           & cuts~(\ref{eq:cuts:pp->bbWj@LHC}) \\\hline
   complete    & 2.5          &    0.34    &    4.6 / 1.4 &  2.6 / 0.8 \\      
   single top  & 0.38         &    0.13    &    1.4 / 0.7 &  0.8 / 0.4 \\\hline\hline
   \multicolumn{1}{|l|}{$g\bar{d}\to \bar{u}W^+b\bar{b}$}
               & soft         & optimized  & soft         & optimized \\
   \multicolumn{1}{|l|}{\quad/ $gd\to uW^-b\bar{b}$}
               & cuts~(\ref{eq:soft-cuts@TeV})
                              & cuts~(\ref{eq:cuts:pp->bbWj@TeV})
                                           & cuts~(\ref{eq:soft-cuts@LHC})
                                           & cuts~(\ref{eq:cuts:pp->bbWj@LHC}) \\\hline
   complete    & 0.41         &    0.08    &   6.0 / 15.2 &  1.7 / 4.0 \\      
   single top  & 0.12         &    0.07    &   4.0 /  9.0 &  1.6 / 3.6 \\\hline
\end{tabular}
\caption{\label{tb:crs1}%
  Single top cross sections at Tevatron and LHC.  The numbers of the
  cuts refer to the formulae in the text.}
\end{center}
\end{table}

The numbers labeled `complete' correspond to the contribution from
all~SM diagrams including the single top single and all
interferences.  The numbers show that the optimized
cuts~(\ref{eq:cuts:optimized}) do indeed improve the signal to
background ratio significantly.  In particular the gluon
initiated subprocesses provide a clean sample that is dominated by
single top production in $W$-gluon fusion.

\begin{figure}
  \begin{center}
    \includegraphics[width=\textwidth]{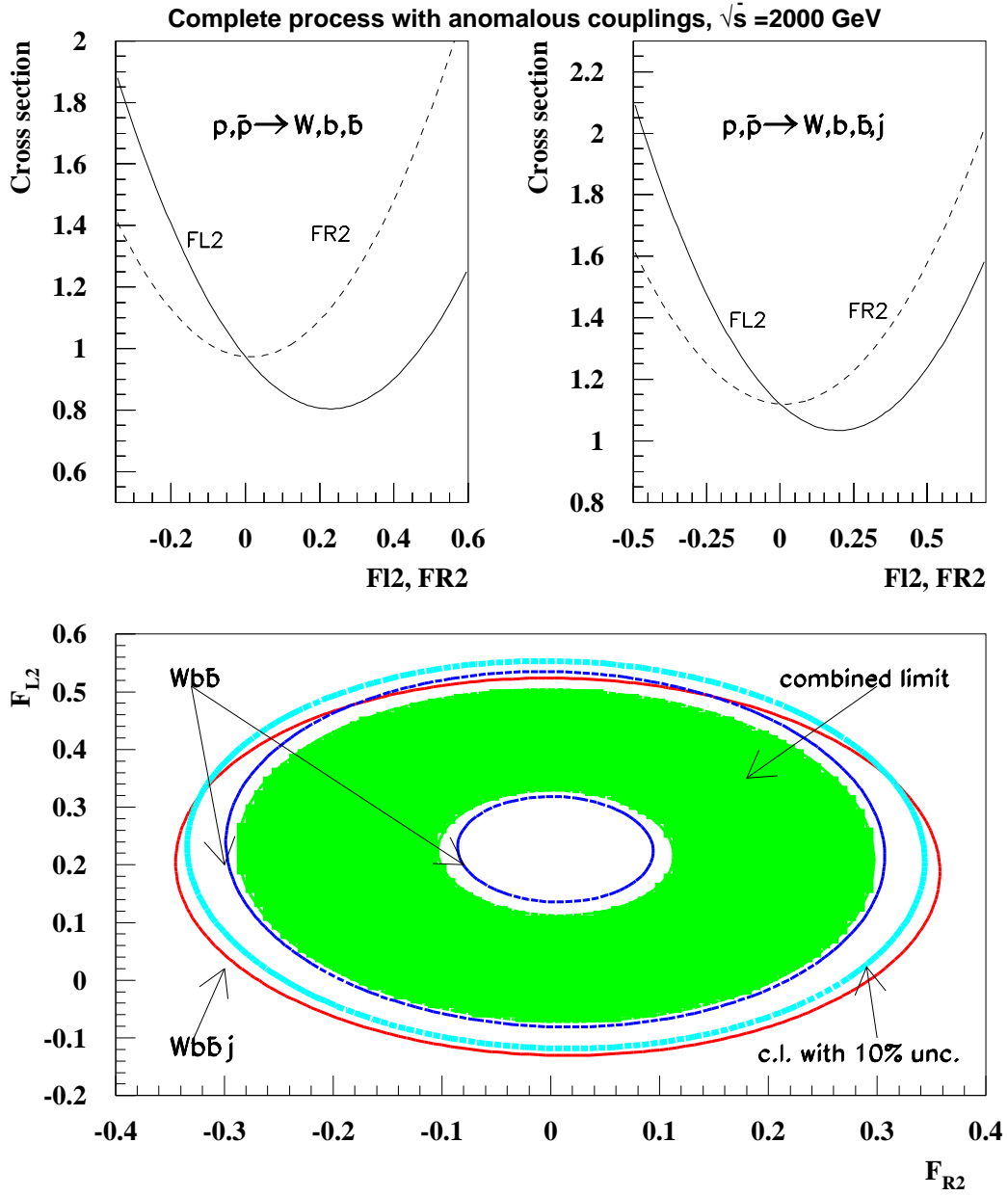}
  \end{center}
  \caption{\label{fig:anom_coup_TEV}%
    Cross sections after optimized
    cuts~(\ref{eq:cuts:pp->bbW@TeV},\ref{eq:cuts:pp->bbWj@TeV})
    and corresponding limits on anomalous couplings at the Tevatron.}
\end{figure}
\begin{figure}
  \begin{center}
    \includegraphics[width=\textwidth]{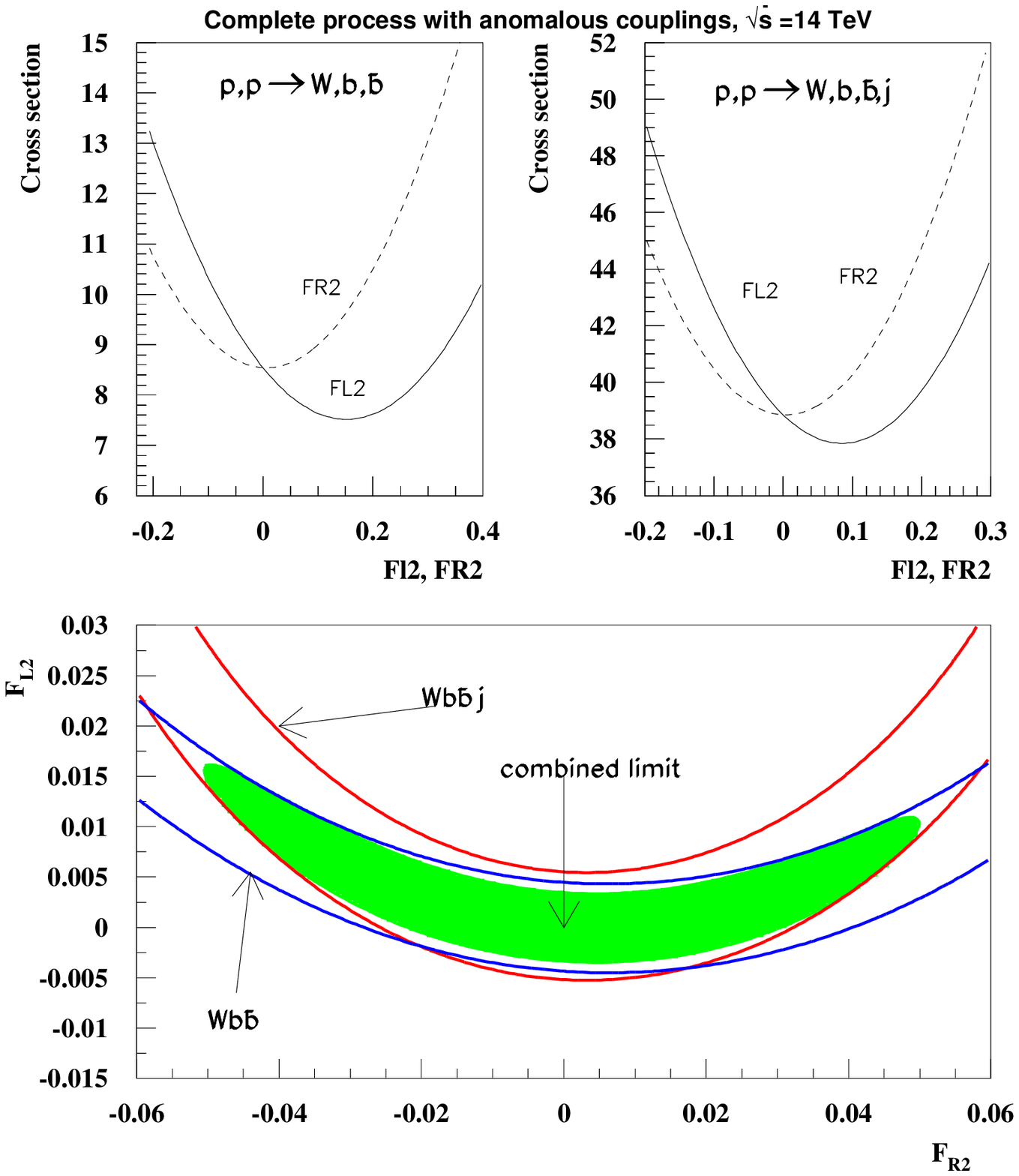}
  \end{center}
  \caption{\label{fig:anom_coup_LHC}%
    Cross sections after optimized
    cuts~(\ref{eq:cuts:pp->bbW@LHC},\ref{eq:cuts:pp->bbWj@LHC})
    and corresponding limits on anomalous couplings at the LHC.}
\end{figure}

The dependence of the total cross section for the
processes~(\ref{eq:pp->bbW(j)}) on anomalous couplings after optimized
cuts~(\ref{eq:cuts:optimized}) is shown
in the upper part of figure~\ref{fig:anom_coup_TEV} for the Tevatron
and in figure~\ref{fig:anom_coup_LHC} for the LHC.  The resulting two
standard deviation exclusion contours are presented in the lower part
of these figures.  These exclusion contours correspond to the
electronic and muonic decay modes of the $W$-boson, including $\tau$
cascade decays.  The combined selection efficiency in the hard
kinematical region under consideration, including the double
$b$-tagging, is assumed to be~$50\%$ and as integrated luminosities we
have used $2\text{fb}^{-1}$ for the upgraded Tevatron and
$100\text{fb}^{-1}$ for the LHC.

The combined annulus in figure~\ref{fig:anom_coup_TEV} corresponds to
the optimistic scenario when only statistical errors are taken into
account.  A systematic uncertainty of about~$10\%$ is expected for the
upgraded Tevatron (cf.~the last paper in~\cite{previous}).  The
resulting exclusion contour is shown in figure~\ref{fig:anom_coup_TEV}
as well (the allowed region now covers the hole of the annulus).

Figure~\ref{fig:anom_coup_LHC} demonstrates that it will be essential
to measure both processes~(\ref{eq:pp->bbW(j)}) at the LHC.  The
allowed regions for the each process alone are rather large annuli,
but the overlapping region is much smaller and allows an improvement
of the sensitivity on anomalous couplings by an order of magnitude
{}from the Tevatron to the LHC.

\begin{figure}
  \begin{center}
    \includegraphics[width=\textwidth]{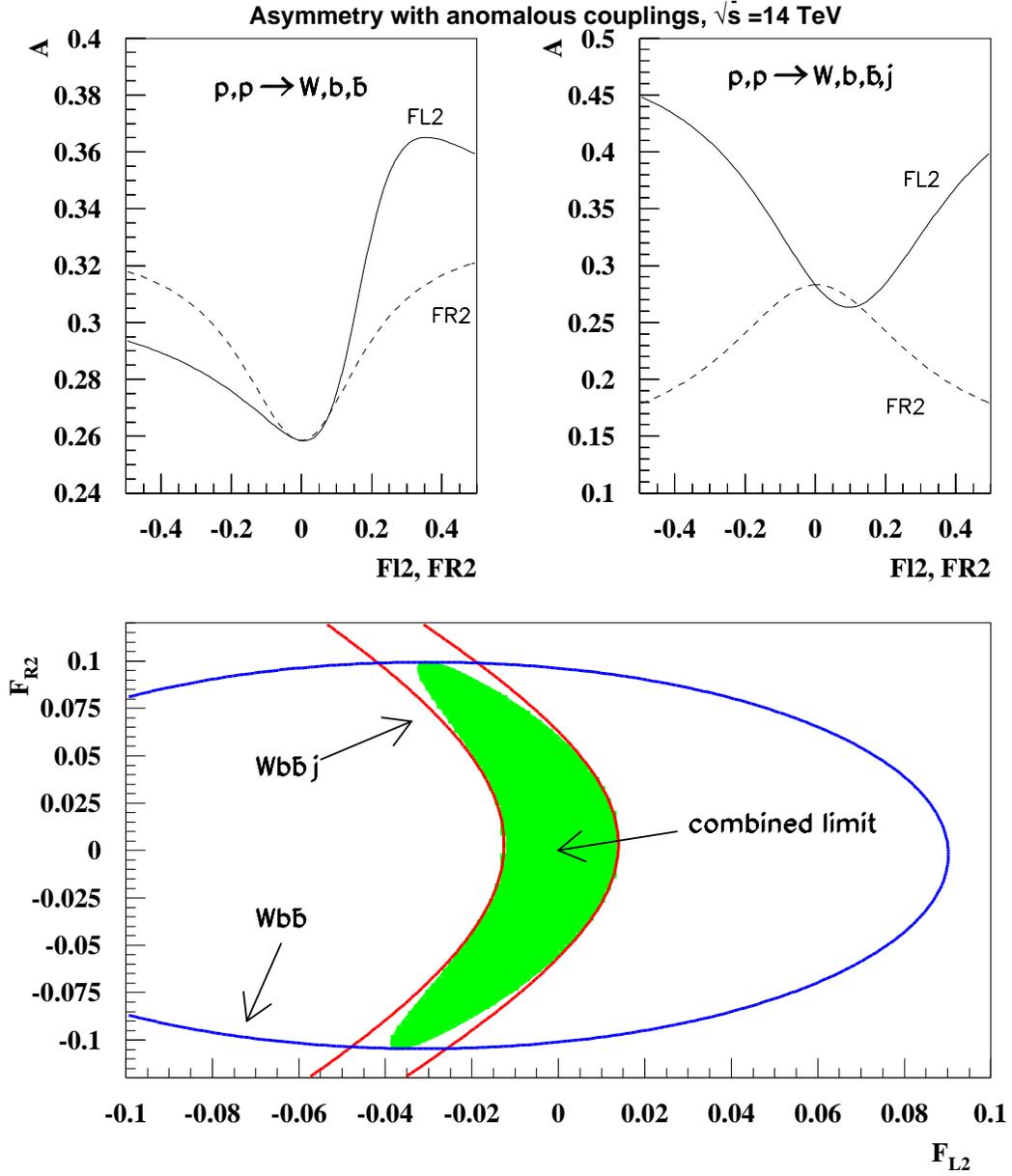}
  \end{center}
  \caption{\label{fig:asymmetry}%
    Top-anti-top asymmetry after optimized
    cuts~(\ref{eq:cuts:pp->bbW@LHC},\ref{eq:cuts:pp->bbWj@LHC})
    and corresponding limits on anomalous couplings at the LHC. (Note
    that the axes are exchanged with respect to
    figures~\ref{fig:anom_coup_TEV} and~\ref{fig:anom_coup_LHC}.)}
\end{figure}
The rate of single top production at LHC is different from the rate of
single anti-top production.  This asymmetry provides an additional
observable at LHC that is not available at the Tevatron.  The
dependence of the asymmetry after optimized
cuts~(\ref{eq:cuts:pp->bbW@LHC},\ref{eq:cuts:pp->bbWj@LHC}) on
anomalous couplings and the resulting two standard deviation exclusion
contours are shown in figure~\ref{fig:asymmetry}.  While the allowed
region for the process~(\ref{eq:pp->bbW}) is different from the region
derived from the rate, the combined limits from both
processes~(\ref{eq:pp->bbW(j)}) are similar.

\begin{figure}
  \begin{center}
    \includegraphics[width=\textwidth]{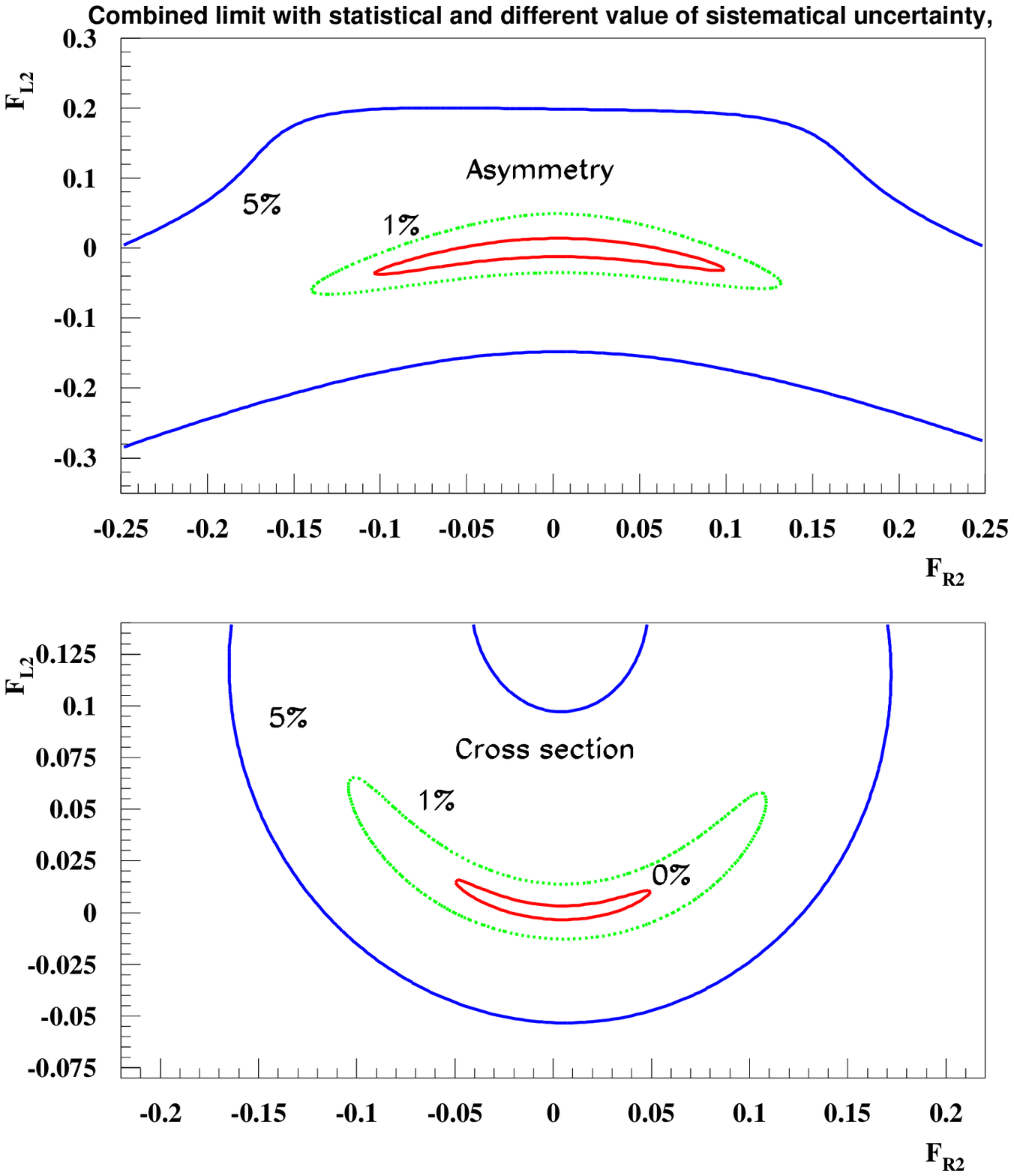}
  \end{center}
  \caption{\label{fig:sistem_unc}%
    Dependence of the limits on anomalous couplings from LHC
    measurements on the size of systematical uncertainties.}
\end{figure}
\begin{table}
  \begin{center}
    \begin{tabular}{|p{2em}rr|lcl|lcl|}\hline
         \multicolumn{3}{|c|}{Systematics}
        &\multicolumn{3}{|c|}{$F_2^L$}
        &\multicolumn{3}{|c|}{$F_2^R$} \\\hline\hline
      &$\pm$&$10\%$ & $-0.18$&$\ldots$&$+0.55$ & $-0.24$&$\ldots$&$+0.25$ \\
      &$\pm$&$ 0\%$ & $-0.07$&$\ldots$&$+0.11$ & $-0.18$&$\ldots$&$+0.21$ \\
      \hline
     \end{tabular}%
   \end{center}%
   \caption{\label{tb:crs2}%
     Uncorrelated limits on anomalous couplings from Tevatron
     measurements with and without systematical uncertainties.}
\end{table}
\begin{table}
  \begin{center}
    \begin{tabular}{|p{2em}rr|lcl|lcl|}\hline
         \multicolumn{3}{|c|}{Systematics}
        &\multicolumn{3}{|c|}{$F_2^L$}
        &\multicolumn{3}{|c|}{$F_2^R$} \\\hline\hline
      &$\pm$&$10\%$ & $-0.094$&$\ldots$&$+0.34$  & $-0.17$ &$\ldots$&$+0.18$ \\
      &$\pm$&$ 5\%$ & $-0.052$&$\ldots$&$+0.097$ & $-0.12$ &$\ldots$&$+0.13$ \\
      &$\pm$&$ 1\%$ & $-0.013$&$\ldots$&$+0.014$ & $-0.05$ &$\ldots$&$+0.06$ \\
      &$\pm$&$ 0\%$ & $-0.003$&$\ldots$&$+0.003$ & $-0.022$&$\ldots$&$+0.03$ \\
      \hline
    \end{tabular}%
  \end{center}%
  \caption{\label{tb:crs3}%
     Uncorrelated limits on anomalous couplings from LHC measurements
     with for several estimates of systematical uncertainties.}
\end{table}

Systematical uncertainties (from $\Delta M_W$, $\Delta M_t$, parton
distribution functions, QCD scales, out of cone corrections, luminosity
determination, etc.) will play an important role at the LHC as well.
It is however impossible to predict them accurately before
cross checks from independent measurements at the LHC scale can be
performed.  Therefore we simply take a set of
combined systematic uncertainty and include it into a new fit for each
value.  Figure~\ref{fig:sistem_unc} shows how the exclusion contours
deteriorate when systematic errors of~$1\%$ and~$5\%$ are included.
In the tables~\ref{tb:crs2} and~\ref{tb:crs3}, for Tevatron and LHC
respectively, the uncorrelated bounds on the anomalous coupling
parameters $F_2^L$ and $F_2^R$ are given, assuming different
systematic uncertainties.  Unfortunately, including a~$10\%$ systematic
error at the LHC will diminish the sensitivity significantly and the
allowed regions will be comparable to those obtained at Tevatron.
 
\begin{table}
  \begin{center}
    \begin{tabular}{|l|lcl|lcl|}\hline
        &\multicolumn{3}{|c|}{$F_2^L$}
        &\multicolumn{3}{|c|}{$F_2^R$} \\\hline\hline
      Tevatron ($\Delta_{\text{sys.}}\approx10\%$)
               & $-0.18$ &$\ldots$&$+0.55$  & $-0.24$ &$\ldots$&$+0.25$ \\
      LHC ($\Delta_{\text{sys.}}\approx5\%$)
               & $-0.052$&$\ldots$&$+0.097$ & $-0.12$ &$\ldots$&$+0.13$ \\
      $\gamma e$ ($\sqrt{s_{e^+e^-}}=0.5\text{TeV}$)
               & $-0.1$  &$\ldots$&$+0.1$   & $-0.1$  &$\ldots$&$+0.1$  \\
      $\gamma e$ ($\sqrt{s_{e^+e^-}}=2.0\text{TeV}$)
               & $-0.008$&$\ldots$&$+0.035$ & $-0.016$&$\ldots$&$+0.016$ \\
      \hline
    \end{tabular}%
  \end{center}%
  \caption{\label{tb:par}%
    Uncorrelated limits on anomalous couplings from measurements at
    different machines.}
\end{table}
The potential of the hadron colliders should be compared to the
potential a next generation $e^+e^-$ linear collider~(LC) where the
best sensitivity could be obtained in high energy $\gamma
e$-collisions~\cite{boos1,young}.  The results of this comparison are
shown in the Table~\ref{tb:par}.  One can see that a
$500\text{GeV}$~LC will outperform the Tevatron (assuming a systematic
uncertainty of $10\%$) by a factor of two to five.  Nevertheless, the
upgraded Tevatron is expected start with physics runs long before a
LC.  The upgraded Tevatron will therefore be able to perform the first
direct measurements of the structure of the $Wtb$~coupling.

The LHC will only be able to rival a $500\text{GeV}$~LC, when the
systematic uncertainties can be kept very small (on the order of
$1\%$).  This goal will be very difficult to achieve.  In the more
realistic scenario of $5\%$ systematic uncertainties, the LHC will
improve the Tevatron limits considerably, but it will fall short of a
high energy LC by a factor of three to eight, depending on the
coupling under consideration.

In the present analysis, we have not included sources
of reducible background~\cite{boos3,willenbrock1} to single top production at
hadron colliders.  However, this reducible background is
sufficiently suppressed in the kinematical regions corresponding to our
optimized cuts.  More detailed simulations and the actual analysis
should nevertheless include the tails of these background
distributions as well.

Finally, we mention that the exclusion contours in
figures~\ref{fig:anom_coup_TEV} and~\ref{fig:sistem_unc} can
can be combined with constraints~\cite{whisnant, young} on the
right-handed coupling $-0.0015 < F^R_2 < 0$  from the CLEO measurement
of~$b \to s \gamma$~\cite{cleo} to improve the limit on the
left-handed coupling~$F_2^L$.
 
%%%%%%%%%%%%%%%%%%%%%%%%%%%%%%%%%%%%%%%%%%%%%%%%%%%%%%%%%%%%%%%%%%%%%%%%
\section{Conclusions}

We have presented the results of a complete tree level calculation of
the processes $pp(\bar{p})\to Wb\bar{b}$ and~$Wb\bar{b}+\text{jet}$,
taking into account the contributions of anomalous $Wtb$~operators.
These final states simultaneously include the single top signal with
subsequent decays and irreducible Standard Model backgrounds.  We have
determined the most sensitive variables from an analysis of the
singularities of the Feynman diagrams in phase space, in order to
achieve optimal background suppression without sacrificing too much of
the signal.

It was shown that the optimized cuts allow to suppress the background
rate drastically and to extract limits on anomalous coupling
parameters.  The accuracy at the LHC is expected to be better by a
factor of two to three compared to the upgraded Tevatron.
Nevertheless, the Tevatron measurements will provide the first direct
information on the structure of the $Wtb$~coupling.  For the higher
accuracy at LHC, it is essential to measure the two final states
separately, to perform efficient double $b$-tagging at high $P^t$ and
to control the systematic uncertainties at a leverl 
betetr than~$10\%$.

At the LHC, one can reduce the dependence of the results on parton
distribution functions, QCD scales, etc.~by using the asymmetry of
single top and single anti-top production.  remains significant after
all cuts.  Reducible backgrounds are expected to be less important in
the phase space region corresponding to the optimized
cuts~\cite{boos3,willenbrock1}.  Nevertheless, a complete simulation
including reducible backgrounds and realistic detector response will
be required for the final experimental analysis.

%%%%%%%%%%%%%%%%%%%%%%%%%%%%%%%%%%%%%%%%%%%%%%%%%%%%%%%%%%%%%%%%%%%%%%%%
\subsection*{Acknowledgments}
We acknowledge discussions with A.~Pukhov, J.~Womersley 
and B.-L.~Young. 
The work has been supported in part by the grant No.~96-02-19773a
of the Russian Foundation of Basic Research, by the Russian Ministry of
Science and Technologies, and by the Sankt-Petersburg Grant Center.
E.\,B. and L.\,D. would like to thank the D0 collaboration and 
E.\,B. is also grateful to the CMS collaboration
for the kind hospitality during visits at FNAL and CERN. 
T.\,O. acknowledges financial support from Bundesministerium f\"ur
Bildung, Wissenschaft, Forschung und Technologie, Germany.

%%%%%%%%%%%%%%%%%%%%%%%%%%%%%%%%%%%%%%%%%%%%%%%%%%%%%%%%%%%%%%%%%%%%%%%%


\begin{thebibliography}{99}
\bibitem{cdfd0}
  F.~Abe \textit{et al.}, (CDF Collaboration),
    Phys.\ Rev.\ Lett.\ \textbf{D74}, 2626 (1995);
  S.~Abachi \textit{et al.}, (D0 Collaboration),
    Phys.\ Rev.\ Lett.\ \textbf{D74}, 2632 (1995);
  P.~Grannis, plenary talk at the International Conference on High
    Energy Physics, Warsaw, 1996.
\bibitem{ew}
  A.~Blondel, plenary talk at the International Conference on High
    Energy Physics, Warsaw, 1996, CERN Report No.~LEPEWWG/96-02.
\bibitem{peccei}
  R.\,D.~Peccei and X.~Zhang, Nucl.\ Phys.\ \textbf{B337}, 269 (1990);
  R.\,D.~Peccei, S.~Peris, and X.~Zhang,
    Nucl.\ Phys.\ \textbf{B349}, 305 (1991).
\bibitem{boos1}
  E.~Boos, A.~Pukhov, M.~Sachwitz, and H.\, J.~Schreiber, 
    Z.\ Phys.\ \textbf{C75}, 237 (1997);
  Phys.\ Lett.\ \textbf{B404}, 119 (1997).
\bibitem{young}
  J.-J.~Cao, J.-X.~Wang, J.-M.~Yang, B.-L.~Young, and X.~Zhang,
    Phys.\ Rev.\ \textbf{D58} 094004 (1998).
\bibitem{previous}
  D.~Dicus and S.~Willenbrock, Phys.\ Rev.\ \textbf{D34}, 155 (1986);
  C.-P.~Yuan, Phys.\ Rev.\ \textbf{D41}, 42 (1990);
  G.\,V.~Jikia and S.\,R.~Slabospitsky,
    Phys.\ Lett.\ \textbf{B295}, 136 (1992);
  R.\,K.~Ellis and S.~Parke, Phys.\ Rev.\ \textbf{D46}, 3785 (1992);
  G.~Bordes and B.~van~Eijk, Z.\ Phys.\ \textbf{C57}, 81 (1993);
  D.\,O.~Carlson and C.-P.~Yuan, Phys.\ Lett.\ \textbf{B306}, 386 (1993);
  G.~Bordes and B.~van~Eijk, Nucl.\ Phys.\ \textbf{B435}, 23 (1995);
  S.~Cortese and R.~Petronzio, Phys.\ Lett.\ \textbf{B253}, 494 (1991);
  D.O.~Carlson, E.~Malkawi, and C.-P.~Yuan,
    Phys.\ Lett.\ \textbf{B337}, 145 (1994);
  T.~Stelzer and S.~Willenbrock, Phys.\ Lett.\ \textbf{B357}, 125 (1995);
  R.~Pittau, Phys.\ Lett.\ \textbf{B386}, 397 (1996);
  M.~Smith and S. Willenbrock, Phys.\ Rev.\ \textbf{D54}, 6696 (1996);
  D.~Atwood, S.~Bar-Shalom, G.~Eilam, and A.~Soni,
    Phys.\ Rev.\  \textbf{D54}, 5412 (1996);
  C.\,S.~Li, R.\,J.~Oakes, and J.\,M.~Yang,
    Phys.\ Rev.\ \textbf{D55}, 1672 (1997);
  C.\,S.~Li, R.\,J.~Oakes, and J.\,M.~Yang,
    Phys.\ Rev.\ \textbf{D55}, 5780 (1997);
  G.~Mahlon, S.~Parke, Phys.\ Rev.\ \textbf{D55}, 7249 (1997);
  A.\,P.~Heinson, A.\,S. Belyaev, and E.\,E. Boos,
    Phys.\ Rev.\  \textbf{D56}, 3114 (1997);
  T.~Stelzer, Z.~Sullivan, and S.~Willenbrock,
    Phys.\ Rev.\  \textbf{D56}, 5919 (1997);
  T.~Tait and C.-P.~Yuan, MSUHEP-71015, hep-ph/9710372;
  D.~Atwood, S.~Bar-Shalom, G.~Eilam, and A.~Soni,
    Phys.\ Rev.\ \textbf{D57}, 2957 (1998);
  T.~Stelzer, Z.~Sullivan, and S.~Willenbrock,
    Phys.\ Rev.\ \textbf{D58} 094021 (1998);
  A.~Belyaev, E.~Boos, and L.~Dudko,
    Phys.\ Rev.\ \textbf{D59}, 075001 (1999), hep-ph/9806332.
\bibitem{willenbrock}
  M.~Smith and S.~Willenbrock, Phys.\ Rev.\ \textbf{D54}, 6696 (1996);
  T.~Stelzer, Z.~Sullivan, and S.~Willenbrock,
    Phys.\ Rev.\  \textbf{D56}, 5919 (1997).
\bibitem{boos2}
  A.\,P.~Heinson, A.\,S.~Belyaev, E.\,E.~Boos,
    Phys.\ Rev.\ \textbf{D56}, 3114 (1997).
\bibitem{boos3}
  A.~Belyaev, E.~Boos, and L.~Dudko,
    Phys.\ Rev.\ \textbf{D59}, 075001 (1999), hep-ph/9806332.
\bibitem{willenbrock1}
  T.~Stelzer, Z.~Sullivan, and S.~Willenbrock,
    Phys.\ Rev.\ \textbf{D58} 09402 (1998).
\bibitem{jikia-boos}
  N.\,V.~Dokholyan and G.\,V.~Jikia,
    Phys.\ Lett.\ \textbf{B336}, 251 (1994);
  E.\,E.~Boos \textit{et al.}, Z.\ Phys.\ \textbf{C70}, 255 (1996).
\bibitem{higgs}
  A.~Stange, W.~Marciano, and S.~Willenbrock,
    Phys.\ Rev.\ \textbf{D50}, 4491 (1994);
  A.~Belyaev, E.\,E.~Boos, and L.~Dudko, 
    Mod.\ Phys.\ Lett.\ \textbf{A10}, 25 (1995).
\bibitem{comphep}
  E.\,E.~Boos, M.\,N.~Dubinin, V.\,A.~Ilyin, A.\,E.~Pukhov,
    and V.I.~Savrin, INP MSU 94-36/358 and SNUTP-94-116, hep-ph/9503280;
  P.~Baikov \textit{et al.}, in Proc.\ of the Xth Int.\ Workshop on High
    Energy Physics and Quantum Field Theory, QFTHEP-95,
    ed.\ by B.~Levtchenko and V.~Savrin, (Moscow, 1995), p.~101.
\bibitem{pukhov}
  V.\,A.~Ilyin, D.\,N.~Kovalenko, and A.\,E.~Pukhov,
    Int.\ J.\ Mod.\ Phys.\ \textbf{C7}, 761 (1996);
  D.\,N.~Kovalenko and A.\,E.~Pukhov,
    Nucl.\ Instrum.\ and Meth.\ \textbf{A389}, 299 (1997).
\bibitem{cteq}
  H.~Lai, J.~Huston, S.~Kuhlmann, F.~Olness, J.~Owens, D.~Soper,
    W.-K.~Tung, and H.~Weerts (CTEQ Collaboration),
    Phys.\ Rev.\ \textbf{D55}, 1280 (1997).
\bibitem{buchmueller-hagiwara-gounaris1}
  W.~Buchm\"uller and  D.~Wyler, Nucl.\ Phys.\ \textbf{B268}, 621 (1986);
  K.~Hagiwara, S.~Ishihara, R.~Szalarski, and D.~Zeppenfeld,
    Phys.\ Rev.\ \textbf{D48}, 2182 (1993);
  K.~Hagiwara, R.~Szalarski, and D.~Zeppenfeld,
    Phys.\ Lett.\ \textbf{B318}, 155 (1993);
  B.~Grzadkowski and J.~Wudka,
    Phys.\ Lett.\ \textbf{B364}, 49 (1995);
  G.\,J.~Gounaris, F.\,M.~Renard, and N.\,D.~Vlachos,
    Nucl.\ Phys.\ \textbf{B459}, 51 (1996).
\bibitem{whisnant}
  K.~Whisnant, J.\,M.~Yang, B.-L.~Young, and X.~Zhang, 
    Phys.\ Rev.\ \textbf{D56}, 467 (1997).
\bibitem{kane}
  G.\,L.~Kane, G.\,A.~Ladinsky, and C.-P.~Yuan, 
    Phys.\ Rev.\  \textbf{D45}, 124 (1992).
\bibitem{boos-dudko}
  E.~Boos and L.~Dudko, in preparation.
\bibitem{pdg}
  C.~Caso \textit{et al.}, Particle Data Group,
    Eur.\ Phys.\ J.\ \textbf{C3}, 1 (1998).
\bibitem{cleo}
  M.~Alam \textit{et al.}, CLEO Collaboration,
    Phys.\ Rev.\ Lett.\ \textbf{74}, 2885 (1995).
\end{thebibliography}
\end{document}